\documentclass[aps,prc,reprint,amsmath,amssymb,nofootinbib,floatfix]{revtex4-2}

\usepackage{dcolumn}
\usepackage{graphicx}
\usepackage{newtxtext,newtxmath}
\usepackage{bm}
\usepackage{amsmath}
\usepackage{mathrsfs}
\usepackage[compat=1.1.0]{tikz-feynhand}
\usepackage{physics}
\usepackage{bxtexlogo}
\usepackage[samesize]{cancel}
\usepackage{ragged2e}
\usepackage[english]{babel}
\addto\captionsenglish{}

\allowdisplaybreaks

\usepackage[hidelinks]{hyperref}
\graphicspath{{figures/}}
\newcommand{\sigmalength}{a^t_{\Sigma N(I=1/2)}}
\newcommand{\sigmathree}{a^t_{\Sigma N(I=3/2)}}
\newcommand{\cut}{0.9\leq\cos\theta_\pi^{\mathrm{cm}}\leq1.0}

\begin{document}

\preprint{APS/123-QED}

\title{\texorpdfstring{The $\Sigma N$ threshold structure in the $K^- d\to\pi^-\Lambda p$ reaction}{The Sigma N threshold structure in the K- d to pi- Lambda p reaction}}

\author{Shunsuke Yasunaga}
\affiliation{Department of Physics, Institute of Science Tokyo, 2-12-1 Ookayama, Meguro, Tokyo 152-8551, Japan}
\author{Daisuke Jido}
\affiliation{Department of Physics, Institute of Science Tokyo, 2-12-1 Ookayama, Meguro, Tokyo 152-8551, Japan}

\begin{abstract}
The $\Lambda p$ invariant-mass spectrum near the two physical $\Sigma N$ thresholds in the $K^-d\to\pi^-\Lambda p$ reaction is studied as a probe of the low-energy $\Sigma N$ interaction. The reaction is formulated up to two-step processes, and the coupled $\Lambda p$, $\Sigma^+n$, and $\Sigma^0p$ channels are described within a constant-$K$-matrix model constrained by spin-triplet scattering lengths. The dependence of the $\Sigma N\to\Lambda p$ conversion amplitude on these scattering lengths is examined, and the relation between its line shape and nearby pole positions is investigated. The resulting invariant-mass spectra are studied at different incident kaon momenta and compared qualitatively with existing experimental data. The effects of the spin-singlet $YN$ amplitude and the deuteron wave function are also examined. The line shape near the $\Sigma N$ thresholds is found to be particularly sensitive to the complex spin-triplet $I=1/2$ $\Sigma N$ scattering length, indicating that measurements of this spectrum can provide useful constraints on the low-energy $\Sigma N$ interaction.
\end{abstract}

\maketitle

\section{Introduction}\label{sec:int}
Despite decades of progress in strangeness physics, the hyperon–nucleon ($YN$) interaction is not yet fully understood. This is partly due to the lack and limited accuracy of direct $YN$ scattering data: because of the short lifetimes of hyperons, direct $YN$ scattering measurements were for many years largely limited to bubble-chamber experiments~\cite{engenmann_1966,alexander_1968,sechi-zorn_1968,eisele_1971,kadyk_1971,hauptman_1977}, and the resulting data were subject to large statistical and systematic uncertainties. More recently, new measurements by the CLAS collaboration~\cite{rowley_2021} and the J-PARC E40 collaboration~\cite{miwa_2021,miwa_2022} have begun to improve this situation. Precise data in the low-momentum region near the $\Lambda N$ and $\Sigma N$ thresholds remain limited. In this threshold region, isospin breaking due to the mass differences between the $\Sigma N$ charge channels can affect the line shape of the $\Lambda p$ elastic cross section near the two $\Sigma N$ thresholds~\cite{sone_2026}. Hypernuclear data provide another constraint on low-energy $YN$ interactions, in particular on the $\Lambda N$--$\Sigma N$ coupling. The observed separation energies are expected to be sensitive to the $\Lambda N$--$\Sigma N$ coupling~\cite{akaishi_2002}.
In addition, femtoscopic analyses of hadron--hadron correlations in high-energy proton--proton and proton--nucleus collisions have been used to constrain low-energy $YN$ interactions~\cite{adamczewski-musch_2016,acharya_2019,acharya_2020}, although the extracted scattering parameters still involve model dependencies and experimental uncertainties.
On the theoretical side, several approaches have been developed. Meson-exchange models have been constructed in Refs.~\cite{rijken_1999,rijken_2010,nagels_2019,holzenkamp_1989,reuber_1994,tominaga_1998,tominaga_2001}, while effective field theories with chiral SU(3) symmetry have been pursued in Refs.~\cite{savage_1996,korpa_2001,polinder_2006,haidenbauer_2013,petschauer_2020,haidenbauer_2020,haidenbauer_2021,li_2016,li_2018,song_2018,ren_2020}. Another line of work is based on quark models~\cite{fujiwara_1996,kohno_2000,fujiwara_2007,garcilazo_2007}. Lattice QCD calculations investigate $YN$ interactions directly from nonperturbative QCD~\cite{beane_2007,inoue_2010,liu_2026}. Although most of these approaches can reproduce existing $YN$ scattering data within uncertainties, even basic quantities such as scattering lengths and effective ranges remain poorly constrained.

An alternative way to investigate the $YN$ interaction is the $K^-d\rightarrow\pi YN$ reaction through the final state interaction. In particular, the $K^-d\rightarrow\pi^-\Lambda p$ reaction has attracted attention, since the $\Sigma N$ threshold appears as a conspicuous structure in the $\Lambda p$ invariant mass spectrum. Theoretical works in the early stages were carried out in Refs.~\cite{karplus_1959,kotani_1959}. Later studies used experimental data for the $\bar{K}N\rightarrow\pi Y$ conversion~\cite{dalitz_1981,dalitz_1982,torres_1986} and employed the Faddeev formalism~\cite{toker_1979,toker_1981}. 
Experimentally, this reaction has been studied with stopped kaons and with in-flight kaons~\cite{cline_1968,alexander_1969,sims_1971,eastwood_1971,braun_1977,pigot_1985}. A systematic analysis of existing experimental data on the $\Lambda p$ invariant-mass spectrum near the $\Sigma N$ thresholds was carried out in Ref.~\cite{machner_2013}, with particular emphasis on the cusp structure. 
The J-PARC E90 experiment is intended to provide new high-statistics data for this reaction~\cite{ichikawa_2022}. Isospin symmetry breaking in the $\Lambda N$ interaction has also been investigated through the $K^-d\to\pi\Lambda N$ reaction~\cite{iizawa_2022,yasunaga_2025}.
The $K^-d\rightarrow\pi\Sigma N$ reaction has also been extensively studied to investigate the $\Lambda(1405)$ resonance by analyzing the $\pi\Sigma$ invariant mass spectrum, both theoretically~\cite{jido_2009,jido_2011,jido_2013,yamagata-sekihara_2013,miyagawa_2012,miyagawa_2018,ohnishi_2016} and experimentally~\cite{aikawa_2023}. In this way, the $K^-d\rightarrow\pi YN$ reaction serves as a versatile probe for studying strangeness physics.

In this work, we calculate the $\Lambda p$ invariant-mass spectrum around the $\Sigma N$ thresholds in the $K^-d\to\pi^-\Lambda p$ reaction and investigate its sensitivity to the low-energy spin-triplet $\Sigma N\to\Lambda p$ conversion, with particular emphasis on the $I=1/2$ $\Sigma N$ interaction. Extending our previous studies \cite{yasunaga_2025,yasunaga_2025a,yasunaga_2026}, we describe the coupled $\Lambda p$, $\Sigma^+n$, and $\Sigma^0p$ channels within a unitarized $K$-matrix model, explicitly taking into account the physical mass differences between the two $\Sigma N$ channels. We examine how the $\Sigma N$ scattering lengths affect the conversion amplitude and its pole structure, and how these features are reflected in the $\Lambda p$ invariant-mass spectrum of the $K^-d\to\pi^-\Lambda p$ reaction. The resulting spectra are compared with existing experimental data, and the effects of the spin-singlet $YN$ interaction and the deuteron wave function are also discussed. Through these analyses, we explore the possibility of constraining the low-energy $\Sigma N$ interaction from the line shape of the $\Lambda p$ invariant-mass spectrum, particularly the threshold structure, in the $K^-d\to\pi^-\Lambda p$ reaction.

The remainder of this article is arranged as follows. Section~\ref{sec:for} presents the formulation of the reaction amplitudes and the coupled-channel $YN$ interaction. Section~\ref{sec:res} presents the parameter dependence of the $\Sigma^+n\rightarrow\Lambda p$ conversion $T$-matrix, its pole structure, and the $\Lambda p$ invariant-mass spectrum, including the incident-momentum dependence and the effects of the two physical $\Sigma N$ thresholds. Section~\ref{sec:dis} compares the spectra with experiment and discusses their sensitivity to the spin-singlet $YN$ amplitude and the deuteron wave function. Finally, Sec.~\ref{sec:sum} is devoted to a summary of this work.

\section{Formulation}\label{sec:for}

\subsection{Kinematics}
The $\Lambda p$ invariant mass spectrum of the $K^-d\rightarrow\pi^-\Lambda p$ reaction is given by
\begin{equation}
\frac{d\sigma}{dM_{\Lambda p}} = \frac{M_d M_\Lambda M_p}{(2\pi)^5 4 k_\text{cm} E_\text{cm}^2}\int d\Omega d\Omega^* |\mathcal{T}|^2|\bm{p}_\pi| |\bm{p}_\Lambda^*|, \label{equ:cross_section}
\end{equation}
where $M_d$, $M_\Lambda$, and $M_p$ are the masses of the deuteron, the $\Lambda$ baryon, and the proton, respectively.

The quantities defined in the overall $K^-d$ center-of-mass (c.m.) frame are the magnitude of the initial-state momentum $k_\text{cm}$, the total energy $E_\text{cm}$, the pion momentum $\bm{p}_\pi$ and the solid angle of the pion $\Omega$. The momentum of the $\Lambda$ baryon $\bm{p}_\Lambda^*$ and its solid angle $\Omega^*$ are defined in the $\Lambda p$ rest frame. The spin-averaged squared amplitude is given by
\begin{equation}
    |\mathcal{T}|^2 = \frac{1}{3} \sum_{a=1}^3 \tr[\mathcal{T}_a \mathcal{T}_a^\dagger].\label{equ:spinsum}
\end{equation}
where the factor $1/3$ averages over the three initial deuteron spin states~$(a=1,2,3)$, while the trace sums over the spin states of the final $\Lambda$ and proton.

We restrict the phase-space integration in Eq.~\eqref{equ:cross_section} to a kinematic region in which the $YN$ final-state interaction is enhanced relative to the background contributions. Specifically, we impose $0.9\leq\cos\theta_\pi^{\mathrm{cm}}\leq1.0$ and $p_p^{\mathrm{lab}}>150$ MeV/$c$, where $\theta_\pi^{\mathrm{cm}}$ is the angle between the incident kaon and the outgoing pion in the overall c.m. frame, and $p_p^{\mathrm{lab}}=|\vec{p}_p|$ is the momentum of the outgoing proton in the deuteron rest frame, which is the laboratory frame in the present setup. These kinematic selections have been shown both experimentally and theoretically to suppress the background contributions~\cite{braun_1977,yasunaga_2025}.

\subsection{Three-body scattering amplitudes}
Unless otherwise stated or indicated by c.m. or an asterisk, the three-momenta and energies entering the reaction amplitudes below are evaluated in the laboratory frame. The phase-space variables $k_\text{cm}$, $E_\text{cm}$, $\bm{p}_\pi$, and $\Omega$ in Eq.~\eqref{equ:cross_section} are defined in the overall c.m. frame; at each phase-space point, the scattering amplitudes are calculated in the laboratory frame using the corresponding final-state momenta.
The elementary meson--baryon amplitudes are originally defined in the corresponding two-body c.m. frames. Following Ref.~\cite{yamagata-sekihara_2013}, we transform them to the laboratory frame and expand the transformed amplitudes in the laboratory scattering angle when constructing the reaction amplitudes below.

\begin{figure}[tbp]
  \centering
  \includegraphics[width=\linewidth]{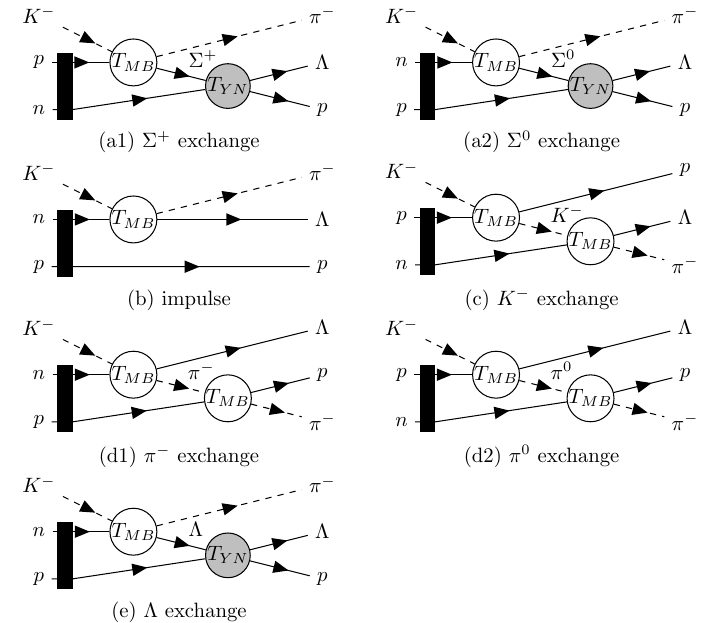}
  \caption{Feynman diagrams for the $K^-d\rightarrow\pi^-\Lambda p$ reaction. The thick line connecting the proton and neutron represents the deuteron, and $T_{MB}$ and $T_{YN}$ denote the meson--baryon and hyperon--nucleon amplitudes, respectively.}
  \label{fig:diagram}
\end{figure}

As shown in Fig.~\ref{fig:diagram}, we consider one- and two-step contributions for the overall amplitude. This truncation is expected to provide a reasonable approximation for deuteron breakup with a fast particle in the final state, for which repeated rescattering involving that particle is expected to be suppressed.
Diagrams~(a1) and (a2) represent the $\Sigma$--exchange contributions of primary interest, consisting of the first-step $K^-N\rightarrow\pi\Sigma$ scattering followed by the second-step $\Sigma N\rightarrow\Lambda p$ conversion. The remaining diagrams describe background processes: diagram~(b) is the one-step impulse process, diagram~(c) is the $K^-$-exchange process, diagrams~(d1) and (d2) are the $\pi^-$- and $\pi^0$-exchange processes, respectively, and diagram~(e) is the $\Lambda$-exchange process.
In the following, we present the explicit amplitudes for these processes. The formulation follows our previous works~\cite{iizawa_2022,yasunaga_2025,yasunaga_2025a} and the detailed treatment originally developed in Ref.~\cite{jido_2009}. The reaction amplitudes are constructed from the meson--baryon amplitudes $T_{MB}$, the hyperon--nucleon amplitudes $T_{YN}$, propagators of intermediate particles, the deuteron spin matrices, and the momentum-space deuteron wave function. In Fig.~\ref{fig:diagram}, the thick line joining the proton and neutron denotes the deuteron, while $T_{MB}$ and $T_{YN}$ label the meson--baryon and hyperon--nucleon amplitudes, respectively. Since we consider spin-averaged observables, the explicit spinors of the final $\Lambda$ and proton are not needed. Instead, the final-state spin sums are evaluated by taking traces over the spin matrices appearing in $\mathcal{T}_a$, as in Eq.~\eqref{equ:spinsum}.

The amplitude of the one-step impulse process represented by diagram~(b) is written as
\begin{equation}
\mathcal{T}^{(I)}_a = T_{K^-n\rightarrow\pi^-\Lambda}(M_{\pi\Lambda},\theta) \sum_{\ell=0,2}\mathcal{S}_a^{(\ell)}(\hat{p}_p)\varphi^{(\ell)}(|\vec{p}_p|),\label{equ:imp}
\end{equation}
where $M_{\pi\Lambda}$ is the invariant mass of the final-state $\pi^-\Lambda$ system, and $\vec{p}_p$ is the momentum of the outgoing proton. A hat denotes the unit vector along the corresponding three-momentum, $\hat p=\vec p/|\vec p|$. The functions $\varphi^{(\ell)}(p)$ are the radial components of the momentum-space deuteron wave function, and $\mathcal{S}_a^{(\ell)}(\hat p)$ are the corresponding spin matrices. The index $\ell$ denotes the internal angular momentum, for which we include the $s$-wave ($\ell=0$) and $d$-wave ($\ell=2$) components. Here and in the hyperon-exchange amplitudes below, $\theta$ is the laboratory angle between the incoming kaon and the outgoing pion, used with the transformed meson--baryon amplitudes.

{
The amplitudes for the hyperon-exchange processes represented by diagrams~(a1), (a2), and (e) are written as
\begin{widetext}
    \begin{equation}
        \begin{aligned}
            \mathcal{T}_a^{(Y)}=&\int\frac{d^3\vec{q}}{(2\pi)^3}G(q;M_Y)\sum_{\ell=0,2}\varphi^{(\ell)}(|\vec{q}+\vec{Q}|)\\
            &\left\{T^1_{YN}(M_{\Lambda p})T_{K^-N\rightarrow\pi^-Y}(W_{K^-N},\theta)\mathcal{S}_a^{(\ell)}(\widehat{q+Q})+T^\sigma_{YN}(M_{\Lambda p})\sum_{b=1}^3\sigma_bT_{K^-N\rightarrow\pi^-Y}(W_{K^-N},\theta)\mathcal{S}_a^{(\ell)}(\widehat{q+Q})\sigma^T_b\right\}.
        \end{aligned}
    \end{equation}
\end{widetext}
Here, $Y$ denotes $\Sigma^+$, $\Sigma^0$, and $\Lambda$ for diagrams~(a1), (a2), and (e), respectively. The quantities $q=(q_0,\vec q)$ and $M_Y$ denote the four-momentum and mass of the exchanged hyperon, and $G(q;M_Y)$ is its propagator, given by
\begin{equation}
    G(q;M_Y) = \frac{2M_Y}{q_0^2-|\vec{q}|^2-M_Y^2+i\varepsilon},
\end{equation}
where the energy component $q_0$ is fixed by energy conservation at the $K^-N\to\pi^-Y$ vertex. With a nonrelativistic simplification for the bound-nucleon energy, we take
\begin{equation}
    q_0^{(Y)} = E_{K^-} + \left(M_N + \frac{|\vec{p}_N|^2}{2M_N}-\frac{B_d}{2}\right) - E_{\pi^-},\label{equ:q0_hyp}
\end{equation}
where $M_N$ and $\vec{p}_N$ are the mass and momentum of the bound nucleon at the first vertex, and $B_d$ is the deuteron binding energy. We include this momentum dependence in the hyperon-exchange amplitudes to reproduce the physical $YN$ threshold positions, in particular the distinct $\Sigma^+n$ and $\Sigma^0p$ thresholds in the $\Lambda p$ invariant-mass spectrum. In each exchange amplitude, $\vec Q$ denotes the momentum transfer at the first vertex, so that $\vec q+\vec Q$ is the momentum of the nucleon inside the deuteron. For hyperon exchange, it is given by
\begin{equation}
    \vec{Q}^{(Y)} = \vec{p}_{\pi^-}-\vec{p}_{K^-}.
\end{equation}
The bound-nucleon momentum in Eq.~\eqref{equ:q0_hyp} is therefore given by $\vec{p}_N=\vec q+\vec Q$. The amplitude $T_{K^-N\to\pi^-Y}(W_{K^-N},\theta)$ describes the first-step meson--baryon process. Here, $N$ denotes a proton for $Y=\Sigma^+$ and a neutron for $Y=\Sigma^0$ or $\Lambda$. The two-body invariant mass entering this amplitude is evaluated by taking the bound nucleon at rest with energy $M_N-B_d/2$:
\begin{equation}
    W_{K^-N} = \sqrt{\left(E_{K^-}+M_N-\frac{B_d}{2}\right)^2-|\vec p_{K^-}|^2}.
\end{equation}

In spin space, we decompose the second-step $s$-wave $YN\to\Lambda p$ amplitude into spin-independent and spin-dependent parts:
\begin{equation}
  T_{YN}(M_{\Lambda p})
  = T^1_{YN}(M_{\Lambda p})
  + T^\sigma_{YN}(M_{\Lambda p})\vec{\sigma}_Y\cdot\vec{\sigma}_N,
\end{equation}
where $M_{\Lambda p}$ is the invariant mass of the final-state $\Lambda p$ system, and $\vec{\sigma}_{Y,N}$ are the Pauli matrices acting on the hyperon and nucleon spin spaces, respectively. In this second-step amplitude, $N$ denotes a neutron for $Y=\Sigma^+$ and a proton for $Y=\Sigma^0$ or $\Lambda$. The index $b$ in the hyperon-exchange amplitudes labels the components of the Pauli matrices, and the superscript $T$ denotes the transpose in baryon spin space. The coefficients $T^1_{YN}$ and $T^\sigma_{YN}$ are related to the spin-singlet and spin-triplet amplitudes, $T^s_{YN}$ and $T^t_{YN}$, by
\begin{equation}
\begin{aligned}
    T^1_{YN} &= \frac{1}{4}\left(T^s_{YN}+3T^t_{YN}\right),\\
    T^\sigma_{YN} &= \frac{1}{4}\left(-T^s_{YN}+T^t_{YN}\right).
\end{aligned}
\end{equation}
As demonstrated in Sec.~\ref{sec:dis}, the spin-singlet $YN$
contribution is relatively small under the kinematic conditions considered
here. We therefore neglect this contribution in the following calculations
unless otherwise stated.

The amplitudes for the meson-exchange processes represented by diagrams~(c), (d1), and (d2) are written as
\begin{widetext}
\begin{equation}
    \mathcal{T}^{(K^-)}_a=\int \frac{d^3\vec{q}}{(2\pi)^3}G(q;M_{K^-})\sum_{\ell=0,2}\varphi^{(\ell)}(|\vec{q}+\vec{Q}|)T_{K^-n\rightarrow\pi^-\Lambda}(M_{\pi\Lambda},\theta_2)\mathcal{S}_a^{(\ell)}(\widehat{q+Q})T^T_{K^-p}(W_{K^-N},\theta_1),
    \label{equ:kex}
\end{equation}
\begin{equation}
    \mathcal{T}^{(\pi)}_a=\int \frac{d^3\vec{q}}{(2\pi)^3}G(q;M_\pi)\sum_{\ell=0,2}\varphi^{(\ell)}(|\vec{q}+\vec{Q}|)T_{\pi N}(M_{\pi^-p},\theta_2)\mathcal{S}_a^{(\ell)}(\widehat{q+Q})T^T_{K^-N\rightarrow\pi\Lambda}(W_{K^-N},\theta_1).
    \label{equ:piex}
\end{equation}
\end{widetext}
In Eqs.~\eqref{equ:kex} and \eqref{equ:piex}, both the $s$- and $d$-wave components of the deuteron are included.
Equation~\eqref{equ:piex} applies separately to the $\pi^-$- and $\pi^0$-exchange contributions. The propagator of the exchanged meson is given by
\begin{equation}
    G(q;M_\text{meson}) = \frac{1}{q_0^2-|\vec{q}|^2-M_\text{meson}^2+i\varepsilon},
\end{equation}
where $q$ is the exchanged-meson four-momentum, and $M_\text{meson}$ is its mass. For pion exchange, $M_\pi$ denotes $M_{\pi^-}$ and $M_{\pi^0}$ for diagrams~(d1) and (d2), respectively. Including the kinetic energy of the bound nucleon, we evaluate $q_0$ at the first vertex as
\begin{align}
    q_0^{(K^-)}&= E_{K^-}+\left(M_p+\frac{|\vec q+\vec Q^{(K^-)}|^2}{2M_p}-\frac{B_d}{2}\right)-E_p,\label{equ:q0_k}\\
    q_0^{(\pi)}&= E_{K^-}+\left(M_N+\frac{|\vec q+\vec Q^{(\pi)}|^2}{2M_N}-\frac{B_d}{2}\right)-E_\Lambda,\label{equ:q0_pi}
\end{align}
where $E_p$ and $E_\Lambda$ are the energies of the outgoing proton and $\Lambda$, respectively, and $M_N$ is the mass of the nucleon at the first vertex. The momentum dependence of $q_0$ is included in the numerical evaluation of the meson-exchange loop integrals. The fixed-energy limit, obtained by omitting the kinetic-energy terms, allows an analytic reduction and is discussed separately in Appendix~\ref{sec:integral}.
The momentum transfers at the first vertex are given by
\begin{align}
    \vec{Q}^{(K^-)}&= \vec{p}_p-\vec{p}_{K^-},\\
    \vec{Q}^{(\pi)}&= \vec{p}_\Lambda-\vec{p}_{K^-}.
\end{align}
For kaon exchange, $T_{K^-p}$ describes the first-step elastic process $K^-p\to K^-p$, and $T_{K^-n\to\pi^-\Lambda}$ describes the second-step production process. For pion exchange, $T_{K^-N\to\pi\Lambda}$ and $T_{\pi N}$ describe the first and second steps, respectively. These processes are $K^-n\to\pi^-\Lambda$ and $\pi^-p\to\pi^-p$ for diagram~(d1), and $K^-p\to\pi^0\Lambda$ and $\pi^0n\to\pi^-p$ for diagram~(d2). The first-step amplitudes use $W_{K^-N}$ defined above, with $M_N=M_p$ for diagrams~(c) and (d2), and $M_N=M_n$ for diagram~(d1). The second-step amplitudes are evaluated at $M_{\pi\Lambda}$ for kaon exchange and $M_{\pi^-p}$ for pion exchange, where $M_{\pi^-p}$ is the invariant mass of the final-state $\pi^-p$ system.
The angles $\theta_1$ and $\theta_2$ are the laboratory scattering angles used with the transformed amplitudes in the first and second steps, respectively. Specifically, $\theta_1$ is the angle between the incoming kaon and the intermediate meson at the first vertex, whereas $\theta_2$ is the angle between the intermediate meson and the outgoing pion at the second vertex.
}

\subsection{Meson--baryon amplitudes}
We employ a partial-wave expansion for the meson--baryon scattering amplitudes. The amplitude $T_{ij}$ for the transition from channel $i$ to channel $j$ is decomposed into non-spin-flip and spin-flip components in baryon spin space as 
\begin{align}
T_{ij} = &\sum_{\ell=0}^\infty \left[(\ell+1)T_{ij}^{\ell+\frac{1}{2}} + \ell T_{ij}^{\ell-\frac{1}{2}}\right]P_\ell(\hat{k}_j\cdot\hat{k}_i) \notag\\
& - i\vec{\sigma}\cdot(\hat{k}_j\times\hat{k}_i)\sum_{\ell=1}^\infty \left(T_{ij}^{\ell+\frac{1}{2}} - T_{ij}^{\ell-\frac{1}{2}}\right)P'_\ell(\hat{k}_j\cdot\hat{k}_i)\label{equ:partial},
\end{align}
where $\ell$ denotes the orbital angular momentum, $P_\ell$ is the Legendre polynomial, and $T_{ij}^{\ell\pm\frac{1}{2}}$ are the partial-wave amplitudes with total angular momentum $j=\ell\pm\frac{1}{2}$. Here, $\hat{k}_i$ and $\hat{k}_j$ are the unit vectors along the meson momenta in the initial and final channels, respectively.
For the $K^-N$ amplitudes, we use the ANL--Osaka DCC model~\cite{kamano_2014}. This model provides partial-wave amplitudes up to $f$-wave ($\ell=3$), which is sufficient for the kinematics considered in this work.
For the $\pi N$ amplitudes, we employ the SAID partial-wave analysis reported in Ref.~\cite{workman_2012}, also including partial waves up to $f$-wave ($\ell=3$), and construct $T_{\pi N}$ using the same convention as in Eq.~\eqref{equ:partial}. Equation~\eqref{equ:partial} gives the original expansion in the corresponding two-body c.m. frame. In the reaction amplitudes, we use the transformed coefficients and laboratory scattering angles described above.

\subsection{Hyperon--nucleon amplitudes}
We describe the three-channel $YN$ system, consisting of $\Lambda p$, $\Sigma^+n$, and $\Sigma^0p$, within a unitarized model based on a $K$-matrix parametrization. The $s$-wave $T$-matrix is given by
\begin{equation}
T_{YN}(W) = -\frac{2\pi W}{\sqrt{M_\Lambda M_\Sigma M_N^2}}(K^{-1}-iP(W))^{-1},\label{equ:kmatrix}
\end{equation}
where $P(W)$ denotes the diagonal matrix of the channel momenta in the ordered particle basis $(\Lambda p,\Sigma^+n,\Sigma^0p)$. In the prefactor of Eq.~\eqref{equ:kmatrix}, we use the averaged masses $M_N=(M_p+M_n)/2$ and $M_\Sigma=(M_{\Sigma^+}+M_{\Sigma^0}+M_{\Sigma^-})/3$ and apply the same mass factor to every matrix element, including the $\Lambda p$ elastic amplitude. This common mass factor is an approximation to channel-dependent baryon mass normalization. The channel momenta in $P(W)$ are evaluated with the physical particle masses. {The ordered isospin basis is given by $(\Lambda p,\Sigma N(I=1/2),\Sigma N(I=3/2))$.} The $K$-matrix in the particle basis is constructed from the $3\times3$ $K$-matrix in the isospin basis through the transformation matrix $C$, which collects the relevant Clebsch--Gordan coefficients:
\begin{align}
    C&=\begin{pmatrix}
        1&0&0\\0&\sqrt{\frac{2}{3}}&\sqrt{\frac{1}{3}}\\0&-\sqrt{\frac{1}{3}}&\sqrt{\frac{2}{3}}
      \end{pmatrix},\\ K_\text{iso}&=\begin{pmatrix}K_{\Lambda N}&K_{\Lambda N-\Sigma N}&0\\K_{\Sigma N-\Lambda N}&K_{\Sigma N(I=1/2)}&0\\0&0&K_{\Sigma N(I=3/2)}\end{pmatrix}.
\end{align}
The $K$-matrix in the particle basis is therefore written as
\begin{equation}
    K = CK_\text{iso}C^T.
\end{equation}
Assuming that the $3\times3$ $K$-matrix is real and symmetric, the model contains four independent real constants: $K_{\Lambda N}$, $K_{\Lambda N-\Sigma N}=K_{\Sigma N-\Lambda N}$, $K_{\Sigma N(I=1/2)}$, and $K_{\Sigma N(I=3/2)}$. These constants are determined by requiring the resulting unitarized $T$-matrix, rather than the $K$-matrix itself, to reproduce the spin-triplet scattering lengths at thresholds. The matching is performed in the isospin basis with the averaged masses $M_N$ and $M_\Sigma$ defined above. The matching channel momenta in $P_\text{iso}$ use these averaged masses, whereas $P$ uses physical particle masses. The conditions are written as
\begin{equation}
\begin{aligned}
    (K_\text{iso}^{-1}-iP_\text{iso}(W))^{-1}_{11}&=-{a^t_{\Lambda p}},\\
    (K_\text{iso}^{-1}-iP_\text{iso}(W))^{-1}_{22}&=-{a^t_{\Sigma N(I=1/2)}}\in \mathbb{C},\\
    (K_\text{iso}^{-1}-iP_\text{iso}(W))^{-1}_{33}&=-{a^t_{\Sigma N(I=3/2)}}.\\
\end{aligned}
\end{equation}
The first condition is evaluated at $W=M_\Lambda+M_N$, and the second and third conditions at $W=M_\Sigma+M_N$, using the averaged masses defined above. Here, the superscript $t$ denotes the spin-triplet channel. The complex $\Sigma N(I=1/2)$ scattering length provides two real constraints through its real and imaginary parts. Thus, the scattering lengths constrain the matrix elements entering the unitarized $T$-matrix and are not assigned directly to the elements of $K_\text{iso}$. For the $\Lambda p$ channel, we consistently use the central value $a^t_{\Lambda p}=-1.56$ fm of the empirical determination $a^t_{\Lambda p}=-1.56^{+0.19}_{-0.22}$ fm~\cite{budzanowski_2010}. The quoted experimental uncertainty is not propagated in the present calculation. For the $\Sigma N$ systems with $I=1/2$ and $I=3/2$, we either scan the scattering lengths or use values predicted by theoretical models.

\subsection{Deuteron wave function and spin matrix}\label{subsec:deuteron}
We employ the momentum-space deuteron wave function obtained from the CD-Bonn potential~\cite{machleidt_2001}. The wave function is defined in the deuteron rest frame, and its $s$-wave ($\ell=0$) and $d$-wave ($\ell=2$) components are parametrized as
\begin{equation}
    \varphi^{(\ell)}(p) = N\sum_i \frac{C_i^{(\ell)}}{p^2+m_i^{(\ell)2}},
\end{equation}
where $p$ is the magnitude of the relative momentum of the two nucleons in the deuteron. The coefficients $C_i^{(\ell)}$ and the mass parameters $m_i^{(\ell)}$ are taken from Table~XX and Eq.~(173), respectively, of Ref.~\cite{machleidt_2001}. The common normalization factor $N$ is fixed using the same momentum-space measure as in the loop integrals:
\begin{equation}
    \int\frac{d^3\vec{p}}{(2\pi)^3}\left[|\varphi^{(0)}(p)|^2+|\varphi^{(2)}(p)|^2\right]=1.
\end{equation}

The spin--angular matrices $\mathcal{S}_a^{(\ell)}$ describe the coupling of a spin-triplet two-nucleon state with $S=1$ and orbital angular momentum $\ell=0,2$ to the total deuteron angular momentum $J=1$. For the $s$-wave, we use
\begin{equation}
    \mathcal{S}_a^{(0)}= S_a^\dag = \frac{i}{\sqrt{2}}\sigma_a\sigma_2.
\end{equation}
For the deuteron $d$-wave, the spin--angular matrix depends on the direction of the relative nucleon momentum $\hat{p}_N$ and is given by
\begin{equation}
    \mathcal{S}_a^{(2)}(\hat{p}_N) =\frac{1}{2\sqrt2}\left(3(\hat{p}_N)_i(\hat{p}_N)_j-\delta_{ij}\right)\sigma_iS^\dag_a(\sigma_j)^T,
\end{equation}
where repeated indices are summed over.

\section{Results}\label{sec:res}
In this section, we examine how the spin-triplet $\Sigma N$ interaction determines the threshold structure in the $K^-d\to\pi^-\Lambda p$ reaction. We first investigate the dependence of the two-body conversion amplitude on the scattering lengths and relate its line shape to nearby poles. We then study how these features appear in the $\Lambda p$ invariant-mass spectrum of the $K^-d\to\pi^-\Lambda p$ reaction for representative interaction parameters.

\subsection{Parameter dependence of the two-body amplitude}\label{sec:param}
\begin{figure*}[t]
 \centering
 \begin{minipage}[t]{0.48\textwidth}\centering
  \includegraphics[width=\linewidth]{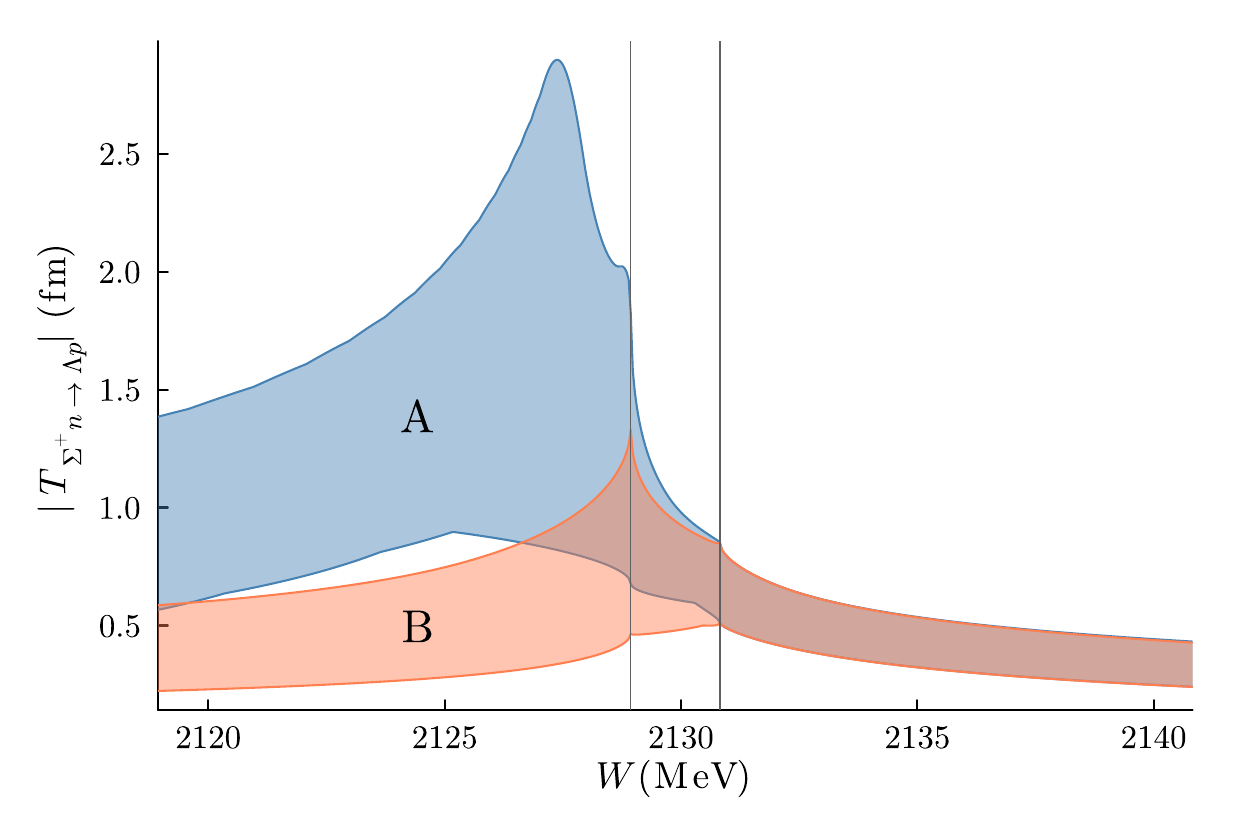}\par
  \vspace{2pt}{\footnotesize(a) Simultaneous variation of the $\Sigma N$ scattering lengths.\par}
 \end{minipage}
 \hfill
 \begin{minipage}[t]{0.48\textwidth}\centering
  \includegraphics[width=\linewidth]{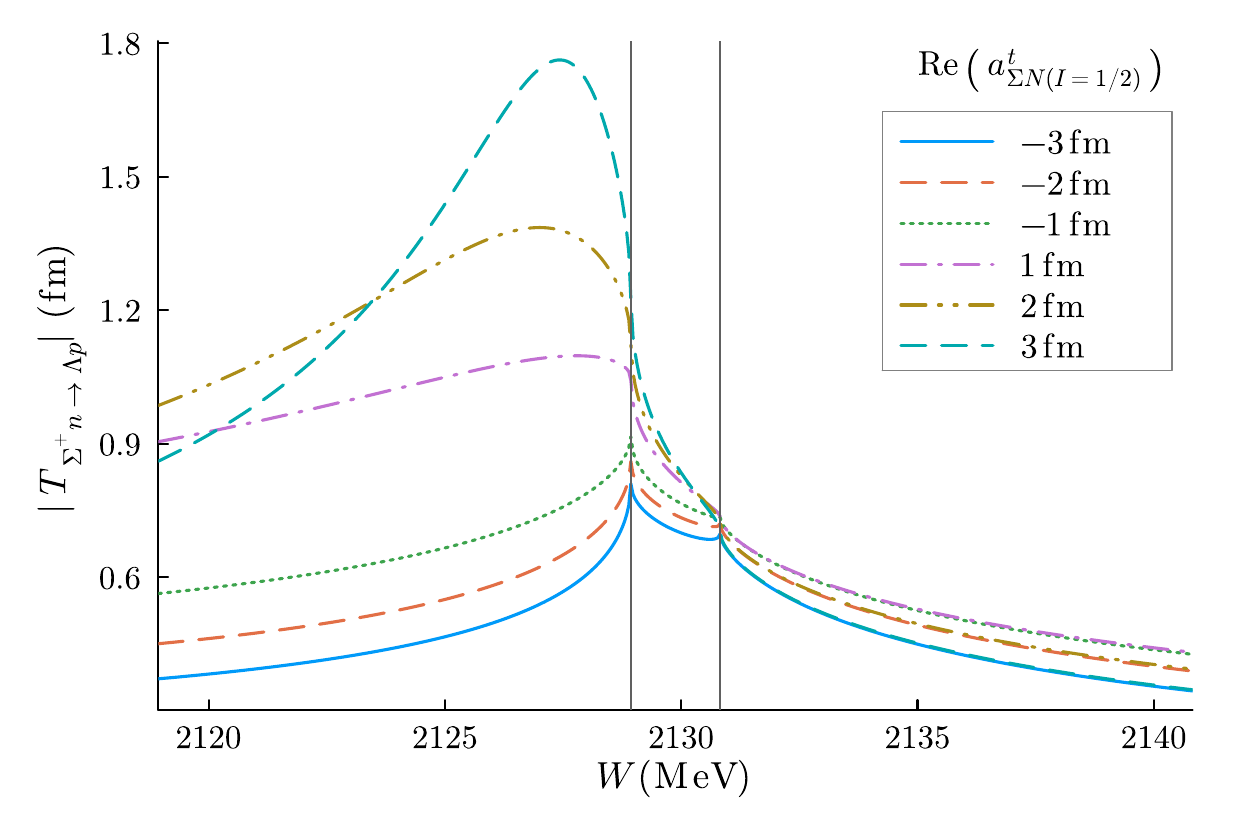}\par
  \vspace{2pt}{\footnotesize(b) Dependence on $\operatorname{Re}\sigmalength$.\par}
 \end{minipage}
 \par\vspace{8pt}
 \begin{minipage}[t]{0.48\textwidth}\centering
  \includegraphics[width=\linewidth]{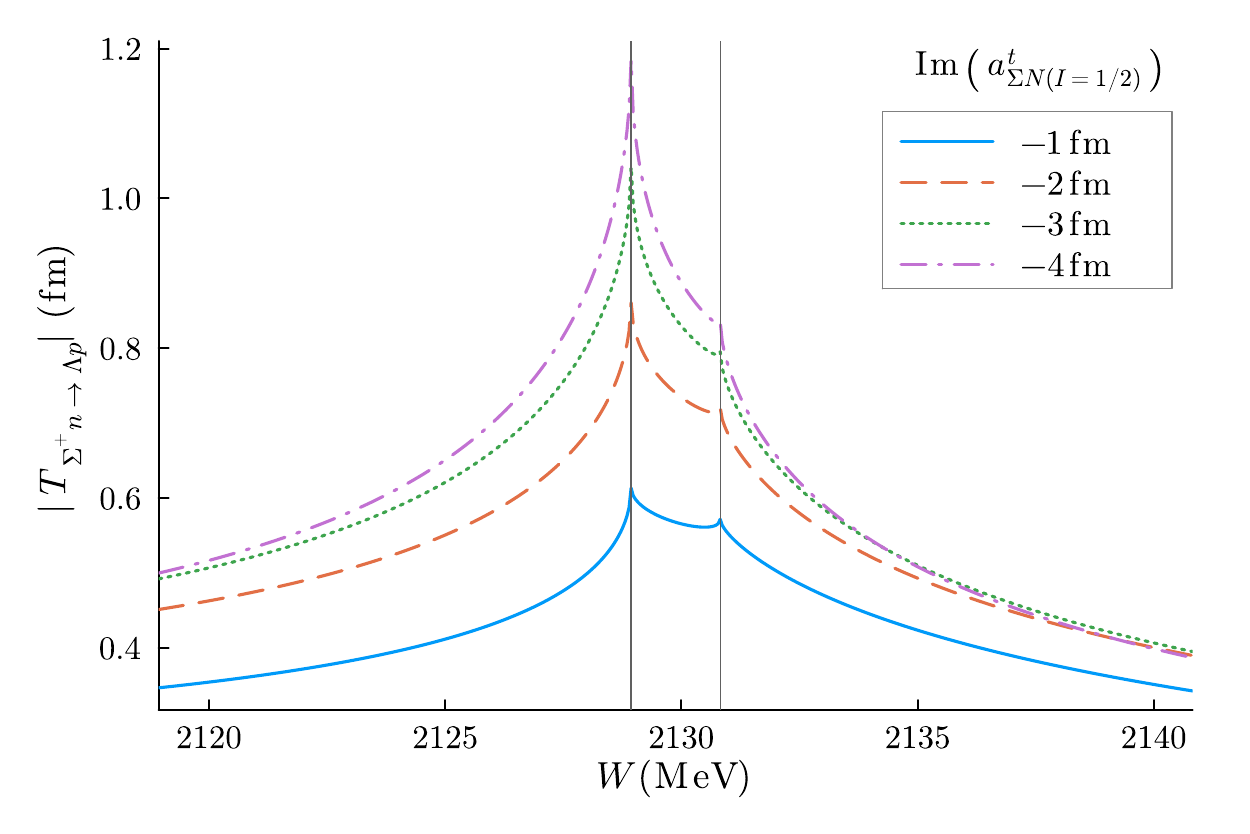}\par
  \vspace{2pt}{\footnotesize(c) Dependence on $\operatorname{Im}\sigmalength$.\par}
 \end{minipage}
 \hfill
 \begin{minipage}[t]{0.48\textwidth}\centering
  \includegraphics[width=\linewidth]{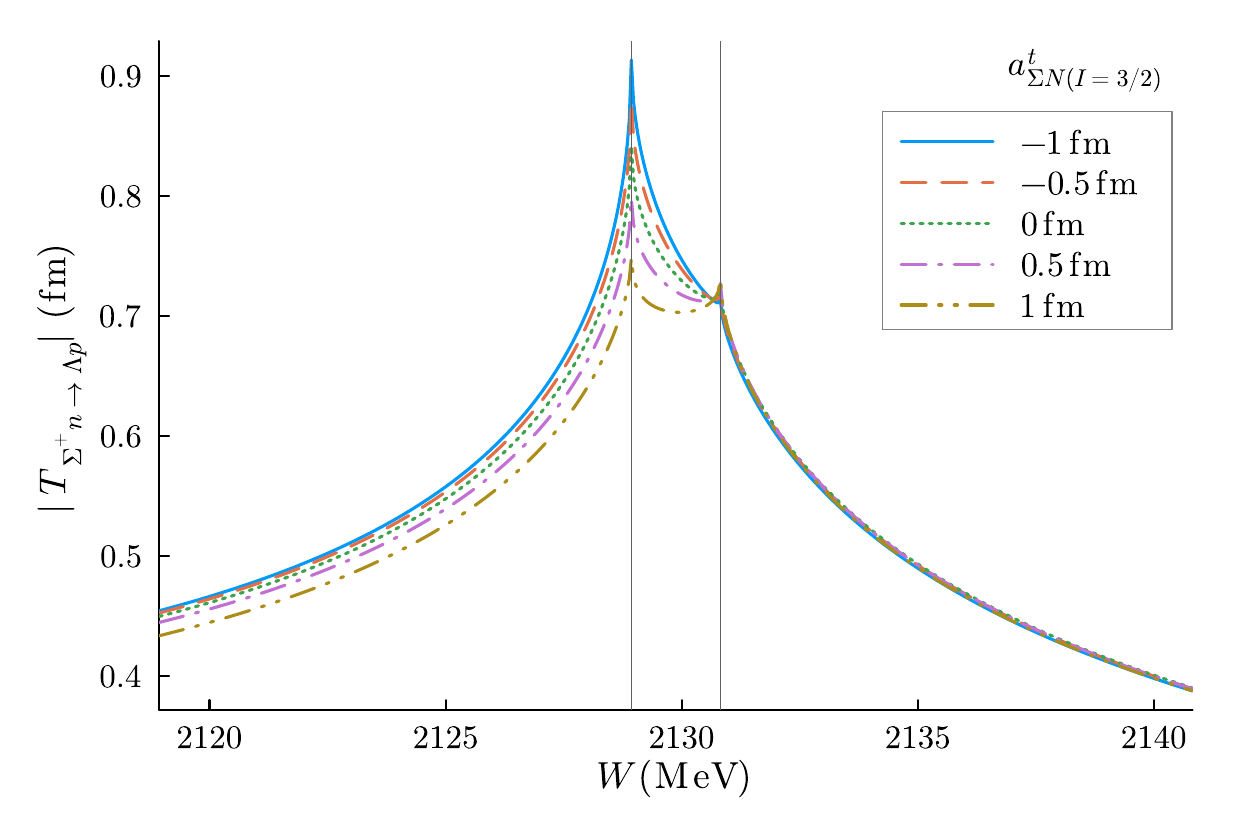}\par
  \vspace{2pt}{\footnotesize(d) Dependence on $\sigmathree$.\par}
 \end{minipage}
 \caption{Magnitude of the spin-triplet $\Sigma^+n\to\Lambda p$ conversion amplitude as a function of the invariant mass $W$. In panel (a), region A and region B show envelopes for $\operatorname{Re}\sigmalength\in[1,4]$ fm and $\operatorname{Re}\sigmalength\in[-4,-1]$ fm, respectively. In both regions, $\operatorname{Im}\sigmalength$ is varied over $[-4,-1]$ fm and $\sigmathree$ over $[-1,1]$ fm.
 Panels (b)--(d) vary one parameter with the others fixed at $\operatorname{Re}\sigmalength=-2$ fm, $\operatorname{Im}\sigmalength=-2$ fm, and $\sigmathree=-0.25$ fm. The $\Lambda p$ scattering length is $a^t_{\Lambda p}=-1.56$ fm. The left and right vertical lines indicate the $\Sigma^+n$ and $\Sigma^0p$ thresholds, respectively.}\label{fig:param}
\end{figure*}

The dependence of the spin-triplet $\Sigma^+n\to\Lambda p$ conversion amplitude on the low-energy $\Sigma N$ parameters is first examined. Figure~\ref{fig:param} shows the magnitude of the conversion amplitude without normalizing factor in Eq.~\eqref{equ:kmatrix} around the physical $\Sigma^+n$ and $\Sigma^0p$ thresholds.
In Fig.~\ref{fig:param}(a), region A is the envelope obtained for $\operatorname{Re}\sigmalength\in[1,4]$ fm, whereas region B is obtained for $\operatorname{Re}\sigmalength\in[-4,-1]$ fm. In both regions, $\operatorname{Im}\sigmalength$ is varied over $[-4,-1]$ fm and $\sigmathree$ over $[-1,1]$ fm. In panels (b)--(d), one parameter is varied at a time while the others are fixed at $\operatorname{Re}\sigmalength=-2$ fm, $\operatorname{Im}\sigmalength=-2$ fm, and $\sigmathree=-0.25$ fm. The vertical lines mark the $\Sigma^+n$ and $\Sigma^0p$ thresholds, in order of increasing invariant mass. The two envelopes exhibit qualitatively different energy dependences. In region A, with $\operatorname{Re}\sigmalength>0$, a broad and potentially large
enhancement develops below the $\Sigma^+n$ threshold. In region B, with $\operatorname{Re}\sigmalength<0$, the amplitude is generally smaller in the subthreshold region and varies more smoothly, apart from the cusp
structures at the two channel openings. Although the sign of the real part of the scattering length cannot be directly identified with an attractive or repulsive interaction in this coupled-channel system, it can significantly affect the line shape of the conversion amplitude. As discussed in Sec.~\ref{sec:pole}, this dependence is closely related to the position of a nearby pole in the complex-energy plane.

The one-parameter variations are displayed in Figs.~\ref{fig:param}(b)--\ref{fig:param}(d). Figure~\ref{fig:param}(b) shows that $\operatorname{Re}\sigmalength$ controls both the magnitude and the qualitative shape of the conversion amplitude, particularly below the lower $\Sigma^+n$ threshold. For negative values of $\operatorname{Re}\sigmalength$, the amplitude increases gradually toward the threshold, whereas positive values produce a broad subthreshold enhancement that becomes more pronounced as $\operatorname{Re}\sigmalength$ increases. This marked change in the subthreshold behavior is closely related to the position of a nearby pole in the complex-energy plane, as discussed in the next section. Figure~\ref{fig:param}(c) shows that making $\operatorname{Im}\sigmalength$ more negative, or equivalently increasing the magnitude of its inelastic part, raises the conversion amplitude over a broad energy interval and accentuates the threshold cusps. By contrast, the dependence on $\sigmathree$ in Fig.~\ref{fig:param}(d) is comparatively weak overall. Its effect is most noticeable near the lower $\Sigma^+n$ threshold. This behavior can be attributed to isospin breaking arising from the finite mass difference between the $\Sigma^+n$ and $\Sigma^0p$ channels. The curves nearly coincide above the $\Sigma^0p$ threshold. These results show that the conversion amplitude is governed primarily by the $I=1/2$ component of the $\Sigma N$ interaction, while the $I=3/2$ component provides a smaller correction in the range considered here.

\subsection{Pole position}\label{sec:pole}
\begin{figure*}[t]
 \centering
 \begin{minipage}[t]{0.48\textwidth}\centering
  \includegraphics[width=\linewidth]{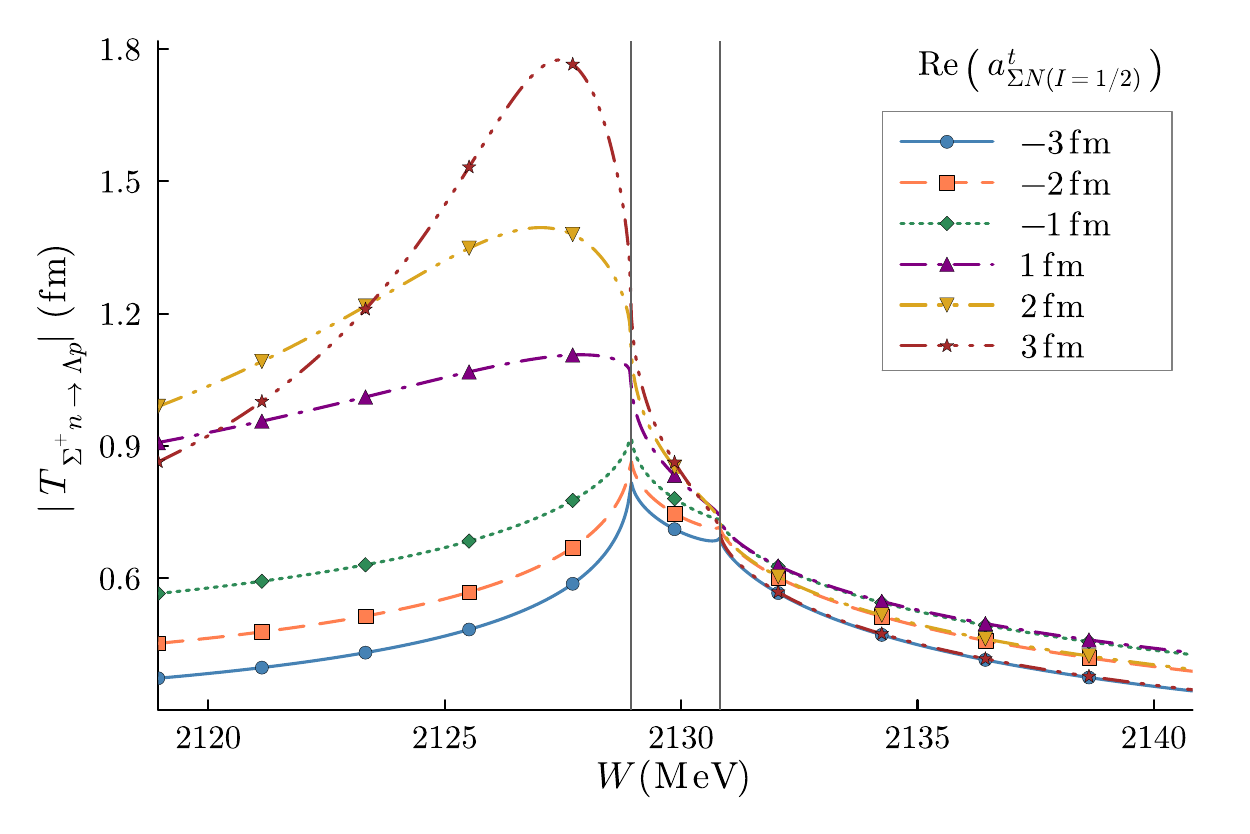}\par
  \vspace{2pt}{\footnotesize(a) Conversion amplitudes for representative scattering lengths.\par}
 \end{minipage}
 \hfill
 \begin{minipage}[t]{0.48\textwidth}\centering
  \includegraphics[width=\linewidth]{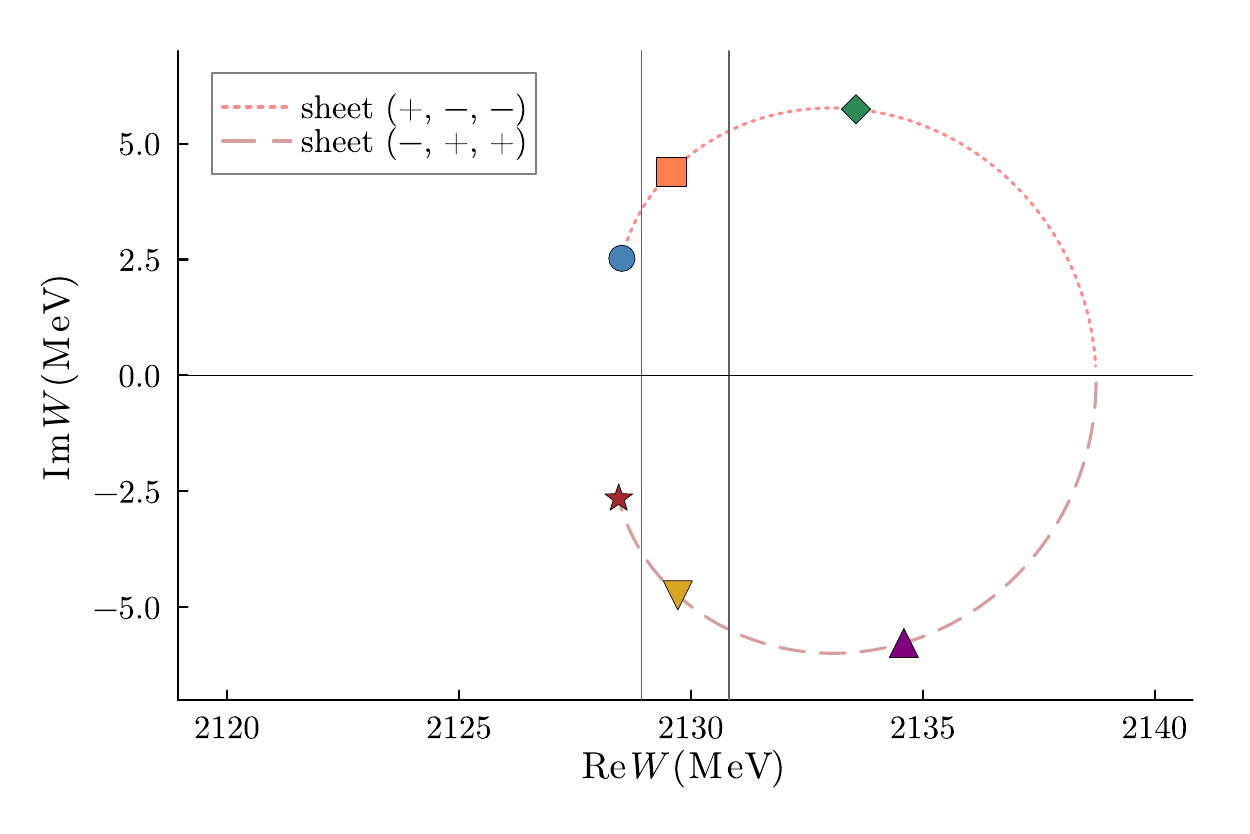}\par
  \vspace{2pt}{\footnotesize(b) Pole trajectories in the complex-energy plane.\par}
 \end{minipage}
 \caption{Conversion amplitudes and nearby poles for
 $\operatorname{Re}\sigmalength$ varied from $-3$ to $3$ fm, with
 $\operatorname{Im}\sigmalength=-2$ fm and $\sigmathree=-0.5$ fm fixed.
 Panel (a) shows representative amplitudes at $-3,-2,-1,1,2,3$ fm. Panel (b) shows the pole trajectories and the
 corresponding representative points. Marker shapes identify the same six parameter values in both panels. Dotted and dashed trajectories denote the $(+,-,-)$ and $(-,+,+)$ sheets, respectively, in channel order $(\Lambda p,\Sigma^+n,\Sigma^0p)$. Only the upper-half-plane branch of $(+,-,-)$ and the lower-half-plane branch of $(-,+,+)$ are shown. The vertical lines indicate the $\Sigma^+n$ and $\Sigma^0p$ thresholds.}\label{fig:pole}
\end{figure*}

To clarify the origin of the qualitatively different threshold structures in Fig.~\ref{fig:param}, the coupled-channel amplitude is analytically continued to the complex-energy plane, and the pole associated with the $\Sigma N$ threshold region is followed. The Riemann sheets are labeled by $(\pm,\pm,\pm)$, where each sign specifies the branch chosen for the corresponding channel momentum in the particle basis $(\Lambda p,\Sigma^+n,\Sigma^0p)$. The physical sheet is defined by $\operatorname{Im}p_i\geq 0$ for each channel momentum $p_i$, corresponding to $(+,+,+)$, while a minus sign indicates the opposite branch.

Figure~\ref{fig:pole}(a) shows the conversion amplitudes for selected values of $\operatorname{Re}\sigmalength$, with $\operatorname{Im}\sigmalength=-2.0$ fm and $\sigmathree=-0.5$ fm fixed, while Fig.~\ref{fig:pole}(b) shows the trajectory of the pole closest to the $\Sigma N$ thresholds. The marker shapes identify the same six scattering-length values in both panels. In panel (b), dotted and dashed lines distinguish the $(+,-,-)$ and $(-,+,+)$ sheets, respectively, in channel order $(\Lambda p,\Sigma^+n,\Sigma^0p)$. The poles on the $(+,-,-)$ and $(-,+,+)$ sheets are shown in the upper and lower half-planes, respectively, while their mirror poles are omitted. The vertical lines indicate the two physical $\Sigma N$ thresholds.

For negative values of $\operatorname{Re}\sigmalength$, the corresponding poles in Fig.~\ref{fig:pole}(b) lie on the $(+,-,-)$ sheet, whose upper-half-plane branch is displayed. The corresponding amplitudes display cusp-like changes at the two thresholds but no pronounced broad enhancement below the $\Sigma^+n$ threshold. For positive values of $\operatorname{Re}\sigmalength$, the relevant poles lie on the lower-half-plane branch of the $(-,+,+)$ sheet and approach the lower threshold and the real axis as this scattering-length parameter increases from 1 to 3 fm.
The enhancement below the $\Sigma^+n$ threshold in Fig.~\ref{fig:pole}(a) becomes progressively larger. This correlation demonstrates that the strong enhancement for positive $\operatorname{Re}\sigmalength$ is associated with the presence and location of a nearby pole rather than with the sign of the scattering length alone. The pole trajectory represents a comparison of different scattering lengths; for each fixed set of values, the pole has a fixed position. Since all of these poles are located on unphysical sheets, their positions must be
interpreted together with the corresponding sheet labels.

\subsection{Invariant mass spectrum}\label{subsec:spectrum}
\begin{figure*}[t]
 \centering
 \begin{minipage}[t]{0.48\textwidth}\centering
  \includegraphics[width=\linewidth]{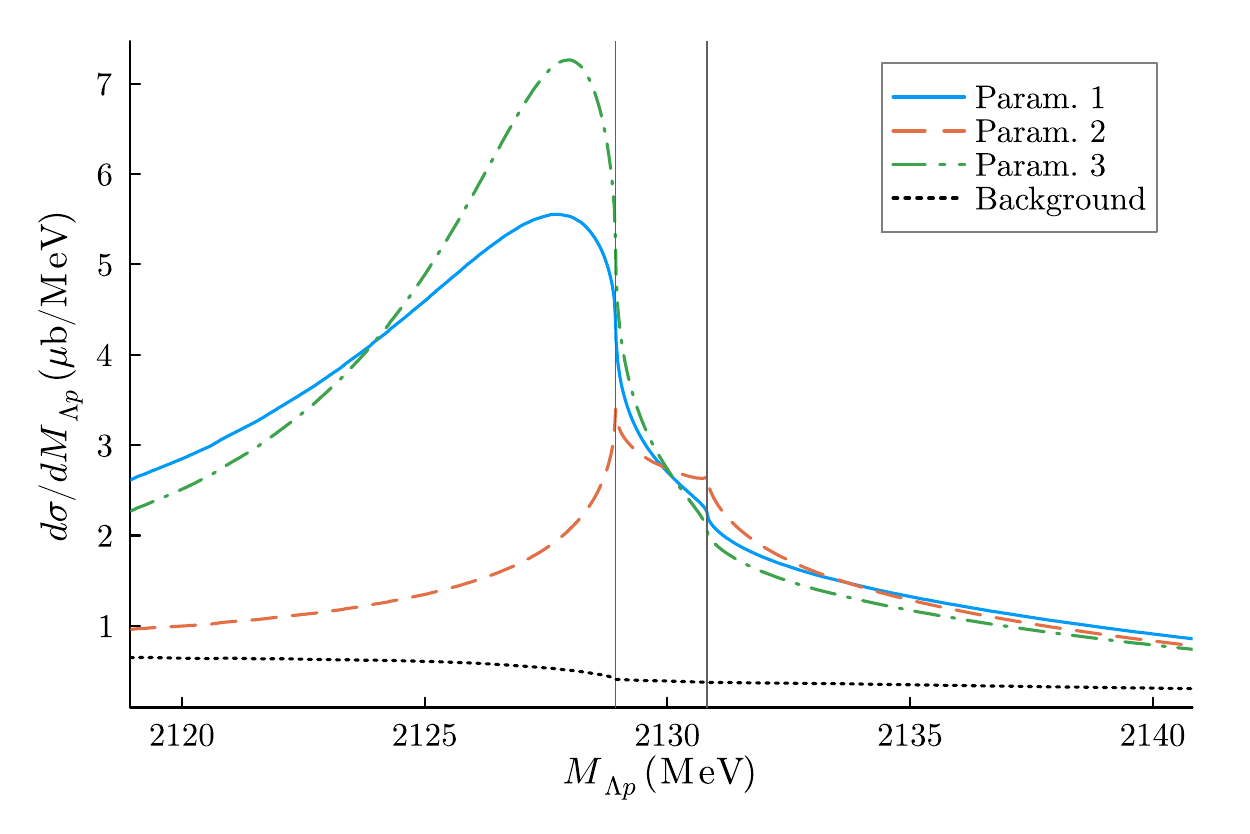}\par
  \vspace{2pt}{\footnotesize(a) $k_{\mathrm{lab}}=1.0$ GeV/$c$.\par}
 \end{minipage}\hfill
 \begin{minipage}[t]{0.48\textwidth}\centering
  \includegraphics[width=\linewidth]{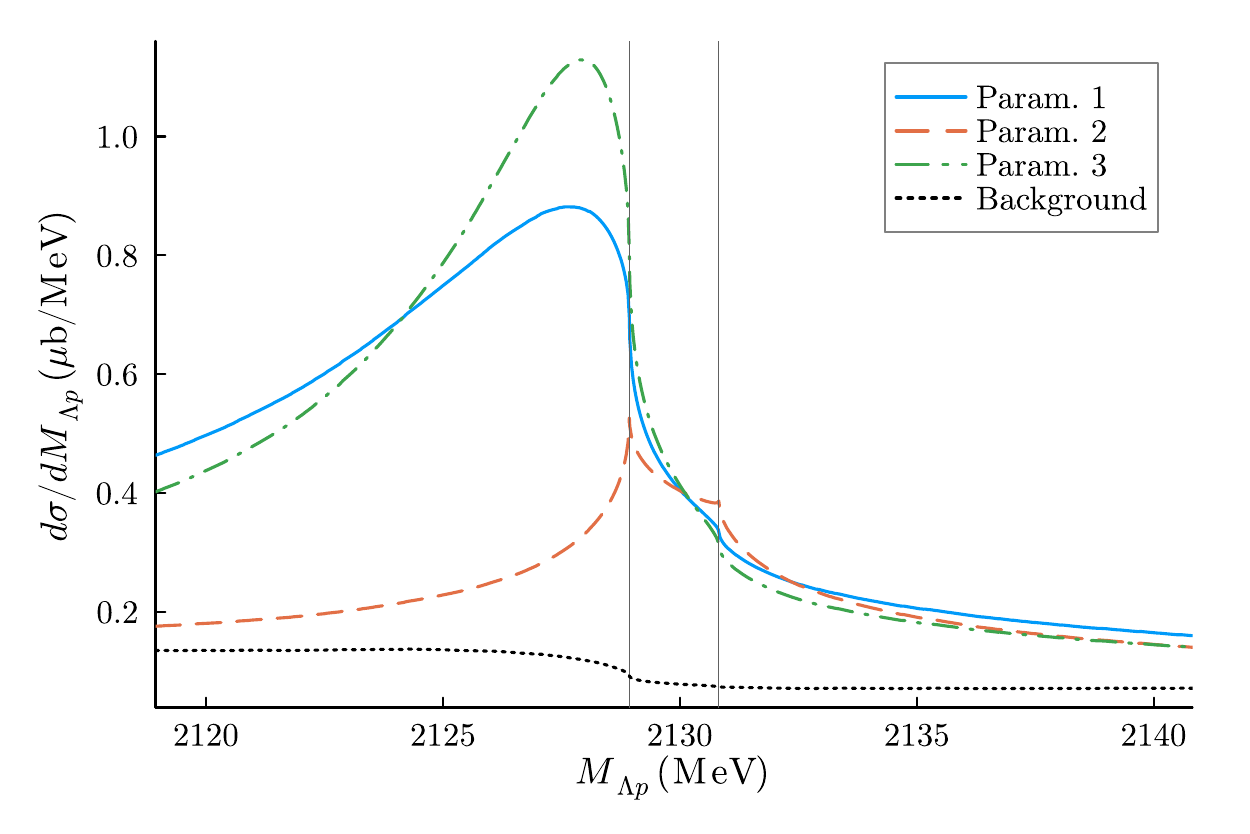}\par
  \vspace{2pt}{\footnotesize(b) $k_{\mathrm{lab}}=1.4$ GeV/$c$.\par}
 \end{minipage}
 \caption{The $\Lambda p$ invariant-mass spectra in the $K^-d\to\pi^-\Lambda p$ reaction at incident kaon momenta of (a) 1.0 and (b) 1.4 GeV/$c$. The solid, dashed, and dash-dotted lines correspond to Param.~1, Param.~2, and Param.~3 in Table~\ref{tab:param}, respectively. The black dotted curves show the background spectra obtained by coherently summing all contributions except $\Sigma$ exchange. The calculations are done with the kinematic selections specified in Sec.~\ref{subsec:spectrum}. The left and right vertical lines indicate the $\Sigma^+n$ and $\Sigma^0p$ thresholds, respectively.}\label{fig:spectrum}
\end{figure*}

We next investigate how the features of the two-body conversion amplitude appear in the $\Lambda p$ invariant-mass spectrum of the $K^-d\to\pi^-\Lambda p$ reaction.
Table~\ref{tab:param} lists three representative sets of spin-triplet $\Sigma N$ scattering lengths taken from existing $YN$ interaction models. In the present constant-$K$-matrix framework, only the quoted scattering lengths determine the model; consequently, the resulting spectra should be regarded as the near-threshold behavior
generated by these low-energy parameters rather than as complete predictions of the original interaction models.

\begin{table}[t]
 \caption{Parameter sets of spin-triplet $\Sigma N$ scattering lengths
 used in the calculation.}
 \label{tab:param}
 \begin{ruledtabular}
 \begin{tabular}{lcc}
 & $\sigmalength$ [fm] & $\sigmathree$ [fm]\\
 Param.~1~\cite{rijken_1999} & $1.68-2.35i$ & $-0.25$\\
 Param.~2~\cite{haidenbauer_2005} & $-3.83-3.01i$ & $0.29$\\
 Param.~3~\cite{haidenbauer_2023} & $2.53-2.64i$ & $0.41$
 \end{tabular}
 \end{ruledtabular}
\end{table}

Figure~\ref{fig:spectrum} shows the spectra calculated with the three parameter sets at incident kaon momenta of 1.0 and 1.4 GeV/$c$. The calculations are performed using the deuteron wave function obtained from the CD-Bonn potential with the kinematic selections $\cut$ and $p_p^{\mathrm{lab}}>150$ MeV/$c$. The solid, dashed, and dash-dotted lines correspond to Param.~1, Param.~2, and Param.~3, respectively. The left and right vertical lines indicate the physical $\Sigma^+n$ and $\Sigma^0p$ thresholds, respectively. The spin-singlet $YN$ amplitude is omitted in these calculations.
For comparison, the dotted curves show the background spectra obtained by coherently summing the impulse, $\Lambda$-, $\bar K$-, and $\pi$-exchange amplitudes, excluding the $\Sigma$-exchange contributions. The background is evaluated under the same conditions as the full spectra, with the $\Lambda N$ amplitude taken from Param.~1 as a common reference. The background spectra are much smaller than the full spectra near the peaks of Param.~1 and Param.~3, indicating the importance of the $\Sigma$-exchange contributions in producing the subthreshold enhancement.

The full spectrum includes interference between $\Sigma$ exchange and the background, so the dotted curve is not an additive contribution to the other curves. The differences among the three parameter sets are most pronounced below the lower threshold and near the two channel openings; the curves approach one another above the
upper threshold. Param.~1 and Param.~3, both having positive $\operatorname{Re}\sigmalength$, exhibit broad enhancements below the lower $\Sigma^+n$ threshold, with Param.~3 producing a larger peak. In contrast, Param.~2, with negative $\operatorname{Re}\sigmalength$, shows a smaller subthreshold enhancement and a sharper cusp at the $\Sigma^+n$ threshold. These differences reflect the dependence of the $\Sigma N\to\Lambda p$ conversion amplitude on the complex $I=1/2$ $\Sigma N$ scattering length, as discussed in Sec.~\ref{sec:param}. As the incident kaon momentum increases from 1.0 to 1.4 GeV/$c$, the overall magnitude of the spectrum decreases, while the characteristic differences among the three parameter sets remain visible.

The physical mass difference between the $\Sigma^+n$ and $\Sigma^0p$ channels leads to two distinct thresholds in the $\Lambda p$ invariant-mass spectrum. The line shape around these thresholds depends on the conversion amplitudes and their interference with the other reaction mechanisms. Therefore, the spectrum near the two $\Sigma N$ thresholds, including the subthreshold region, provides a sensitive probe of the spin-triplet $I=1/2$ $\Sigma N$ interaction.

\section{Discussion}\label{sec:dis}
In this section, we compare the calculated $\Lambda p$ invariant-mass spectra with existing experimental data. We then examine the effects of the spin-singlet $YN$ amplitude and the deuteron wave function on the $\Sigma N$ threshold structure.

\subsection{Comparison with experimental data}
\begin{figure*}[t]
 \centering
 \begin{minipage}[t]{0.48\textwidth}\centering
  \includegraphics[width=\linewidth]{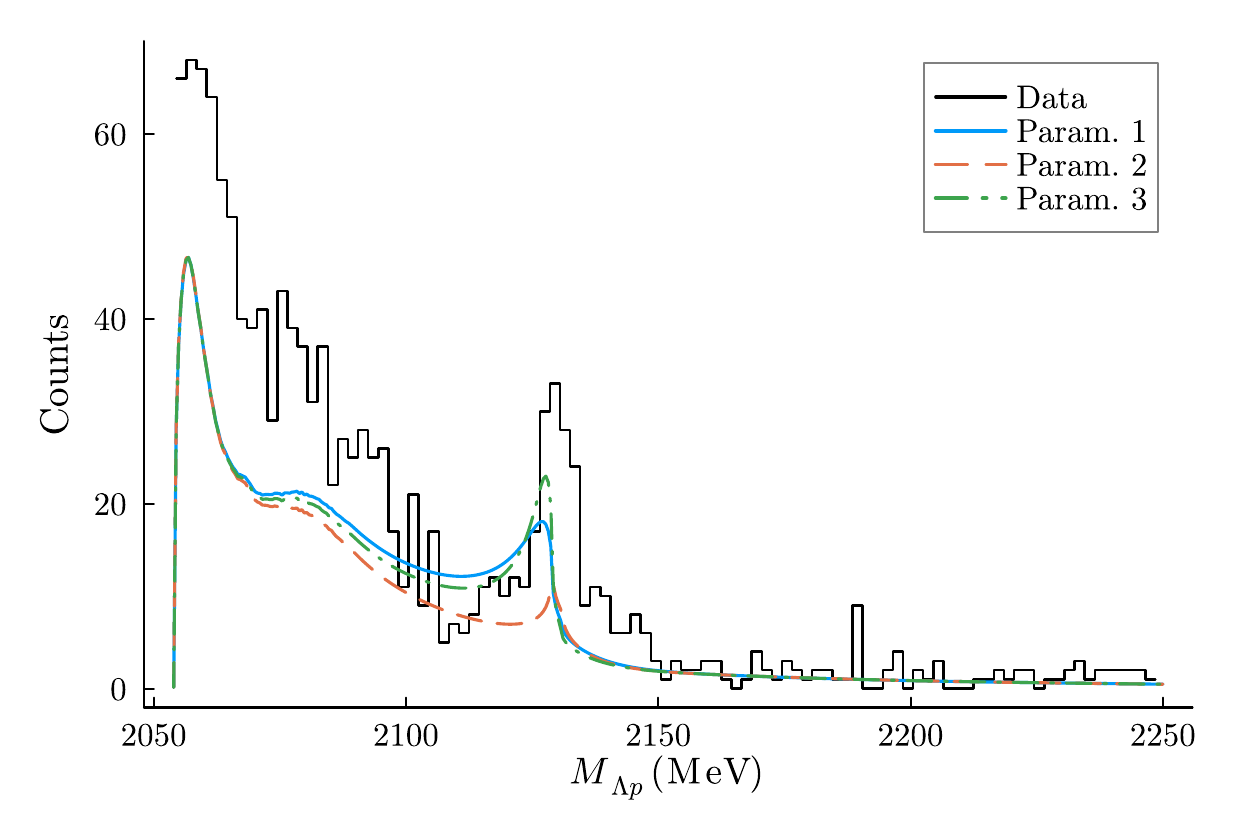}\par
  \vspace{2pt}{\footnotesize(a) $p_p^{\mathrm{lab}}>75$ MeV/$c$.\par}
 \end{minipage}\hfill
 \begin{minipage}[t]{0.48\textwidth}\centering
  \includegraphics[width=\linewidth]{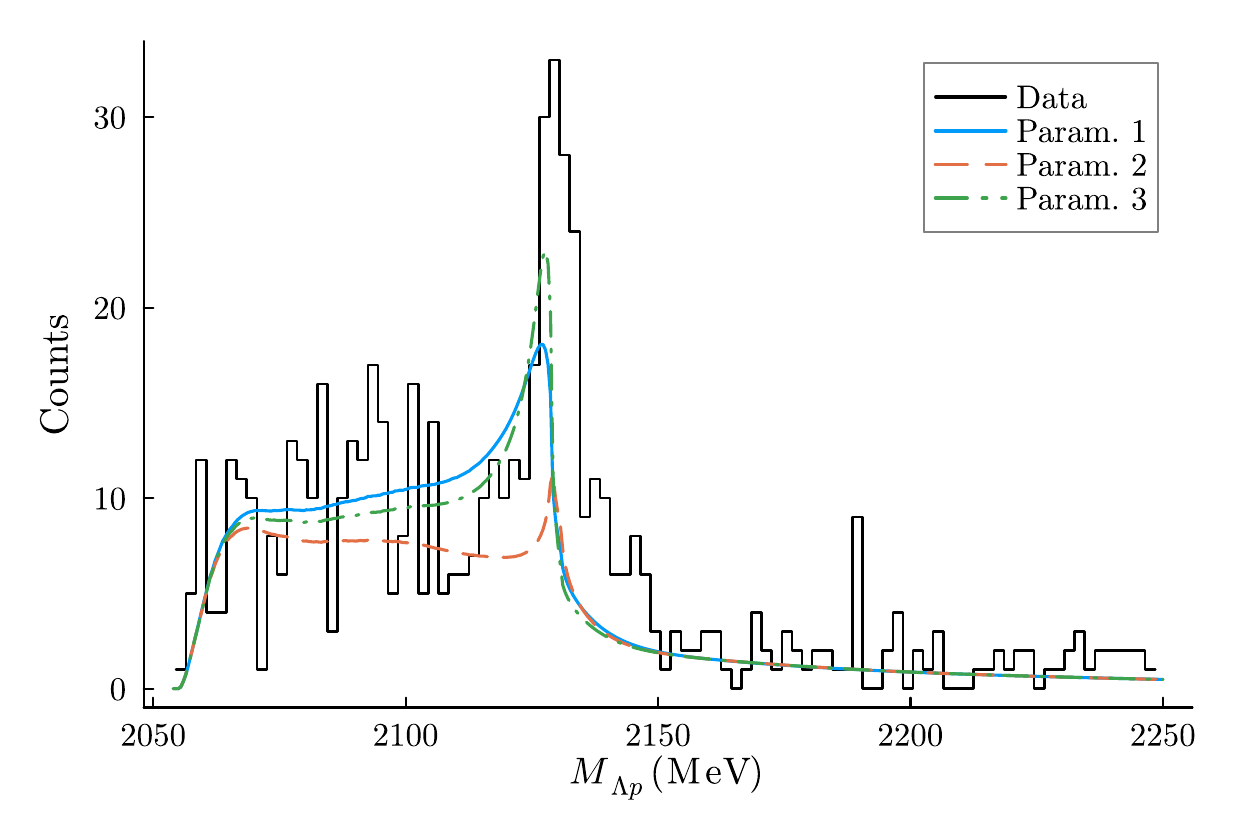}\par
  \vspace{2pt}{\footnotesize(b) $p_p^{\mathrm{lab}}>150$ MeV/$c$.\par}
 \end{minipage}
 \caption{Comparison of the $\Lambda p$ invariant-mass spectra with experimental data~\cite{braun_1977} for proton momenta above (a) 75 and (b) 150 MeV/$c$.
 The histograms show the measured counts in 2 MeV mass bins, and the solid, dashed, and dash-dotted curves correspond to Param.~1, Param.~2, and Param.~3, respectively.
 The calculations use $k_{\mathrm{lab}}=760$ MeV/$c$, the deuteron wave function obtained from the CD-Bonn potential, and the angular selection $\cut$.
 The spin-singlet $YN$ amplitude is omitted.
 The theoretical curves show the differential spectra multiplied by the experimental normalization factor of $7.1$ counts/$\mu$b~\cite{braun_1977}, without fitting. The experimental 2 MeV bin width is not taken into account in the theoretical curves, so the comparison is qualitative.}\label{fig:experiment}
\end{figure*}

The data of Ref.~\cite{braun_1977} were obtained with incident kaon momenta between 680 and 840 MeV/$c$. We perform the calculation at the central momentum of 760 MeV/$c$, following the comparison procedure of Ref.~\cite{yasunaga_2025}. Figure~\ref{fig:experiment} compares the calculated spectra with these data for the proton-momentum selections $p_p^{\mathrm{lab}}>75$ and $p_p^{\mathrm{lab}}>150$ MeV/$c$ in panels (a) and (b), respectively. The calculations use the deuteron wave function obtained from the CD-Bonn potential and the angular selection $\cut$, with the spin-singlet $YN$ amplitude omitted. The histograms show the experimental counts, and the solid, dashed, and dash-dotted curves correspond to Param.~1, Param.~2, and Param.~3, respectively. The theoretical curves are obtained by multiplying the differential spectra by the experimental factor of $7.1$ counts/$\mu$b from Ref.~\cite{braun_1977}, as in Ref.~\cite{yasunaga_2025}. The same factor is applied to all parameter sets and both proton-momentum selections, without further adjustment. The experimental 2 MeV bin width is not taken into account: the theoretical spectra are neither integrated over the bins nor multiplied by the bin width. The comparison therefore concerns the qualitative line shape rather than the absolute counts.

Raising the proton-momentum cut from 75 to 150 MeV/$c$ strongly suppresses the calculated spectrum near the $\Lambda p$ threshold. The spectra around the $\Sigma N$ thresholds remain nearly unchanged, and the differences among the three parameter sets remain visible. Param.~1 and Param.~3 produce broader enhancements below the $\Sigma^+n$ threshold, whereas Param.~2 gives a more localized cusp.

The suppression near the $\Lambda p$ threshold is also seen in the experimental distributions, and the calculation qualitatively reproduces this change with the proton-momentum cut. The present comparison, made without fitting the normalization or the scattering lengths, does not establish a preferred parameter set. More precise measurements of the line shape in this region would help distinguish the predicted spectra and constrain the $\Sigma N$ scattering lengths.

\subsection{Spin-singlet contribution}
\begin{figure}[t]
 \centering\includegraphics[width=\linewidth]{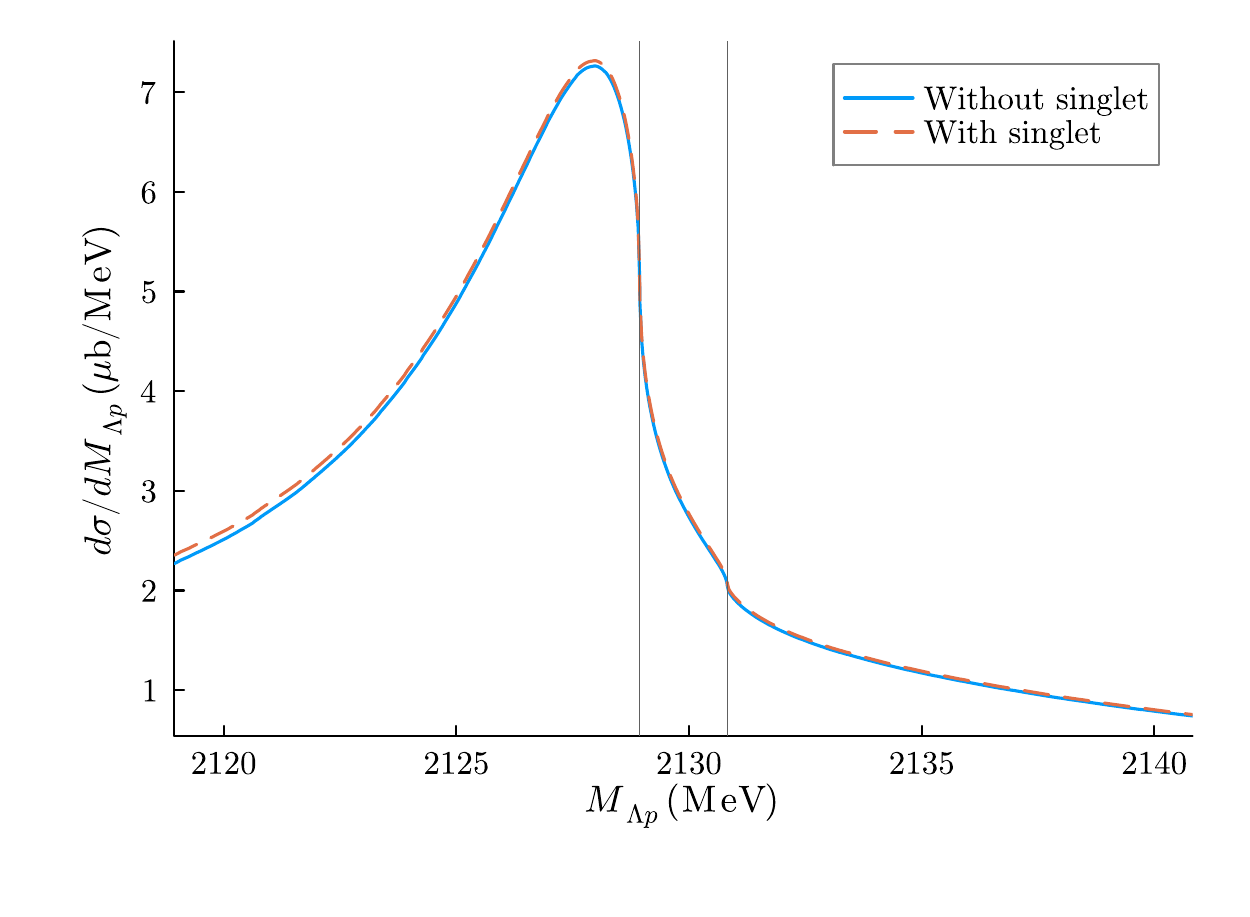}
 \caption{Effect of the spin-singlet $YN$ amplitude on the $\Lambda p$ invariant-mass spectrum.
 The solid and dashed curves show the results without and with the singlet amplitude, respectively.
 The spin-triplet scattering lengths are those of Param.~3, and the spin-singlet scattering lengths are specified in the text.
 The calculations use the deuteron wave function obtained from the CD-Bonn potential, $k_{\mathrm{lab}}=1.0$ GeV/$c$, and the kinematic selections $\cut$ and $p_p^{\mathrm{lab}}>150$ MeV/$c$.
 The left and right vertical lines indicate the $\Sigma^+n$ and $\Sigma^0p$ thresholds, respectively.}\label{fig:singlet}
\end{figure}

Figure~\ref{fig:singlet} compares the spectra calculated with and without the spin-singlet $YN$ amplitude at $k_{\mathrm{lab}}=1.0$ GeV/$c$. The calculations use the deuteron wave function obtained from the CD-Bonn potential and the spin-triplet scattering lengths of Param.~3. The kinematic selections are $\cut$ and $p_p^{\mathrm{lab}}>150$ MeV/$c$, as in Fig.~\ref{fig:spectrum}. The result without the singlet amplitude therefore corresponds to the Param.~3 calculation in Fig.~\ref{fig:spectrum}(a). The spin-singlet scattering lengths are
$a^s_{\Lambda p}=-2.43$ fm,
$a^s_{\Sigma N(I=1/2)}=1.0-1.0i$ fm, and
$a^s_{\Sigma N(I=3/2)}=1.0$ fm.

The solid and dashed curves show the results without and with the singlet amplitude, respectively. Including this amplitude slightly increases the spectrum, with the difference most visible on the low-mass side of the enhancement. The curves nearly coincide around the peak, and the peak position and the line shape near the two $\Sigma N$ thresholds remain essentially unchanged. For the scattering lengths and kinematic selections considered here, the spin-singlet contribution thus gives only a small correction to the threshold structure. This result is consistent with the suppression of the singlet contribution by the forward-pion selection discussed in Ref.~\cite{yasunaga_2025}.

\subsection{Dependence on the deuteron wave function}
\begin{figure*}[t]
 \centering
 \begin{minipage}[t]{0.37816\textwidth}\centering
  \includegraphics[width=\linewidth,trim=0 55 0 20,clip]{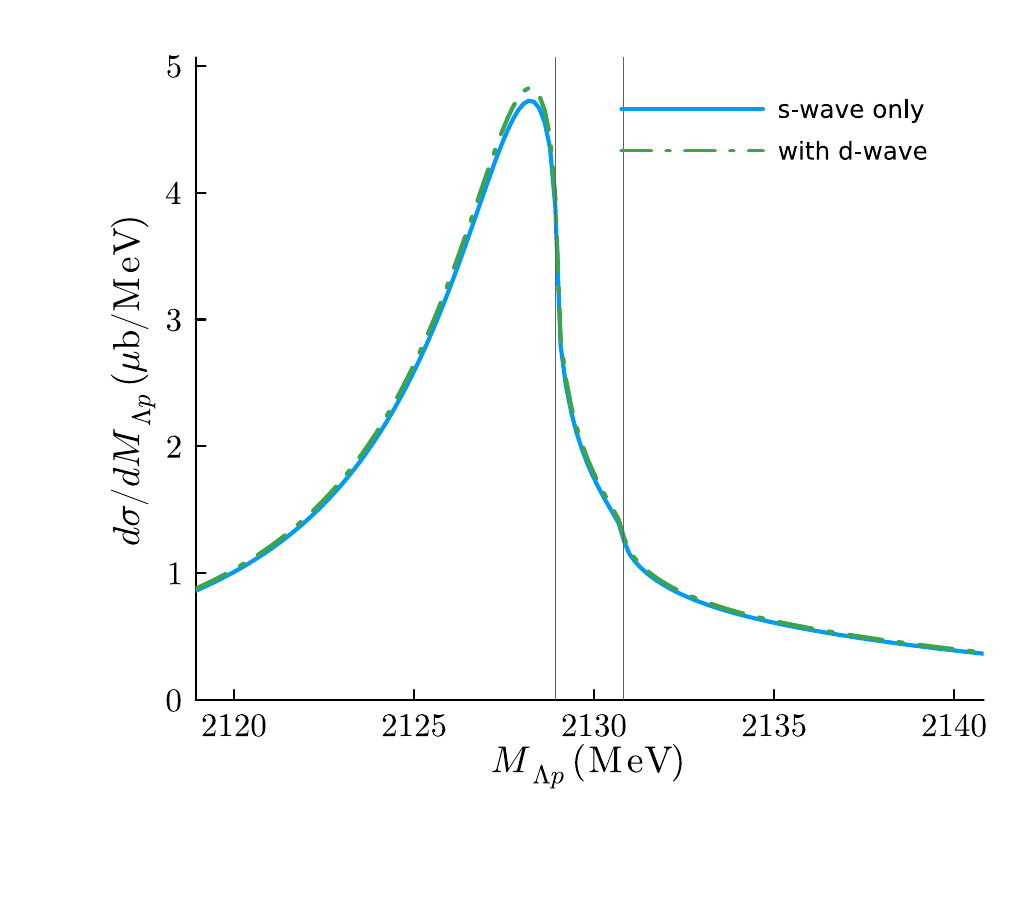}\par
  \vspace{1pt}{\footnotesize
   \hspace*{0.17781\linewidth}\makebox[0.82219\linewidth]{(a) CD-Bonn}\par}
 \end{minipage}%
 \begin{minipage}[t]{0.31092\textwidth}\centering
  \includegraphics[width=\linewidth,trim=86.06 55 0 20,clip]{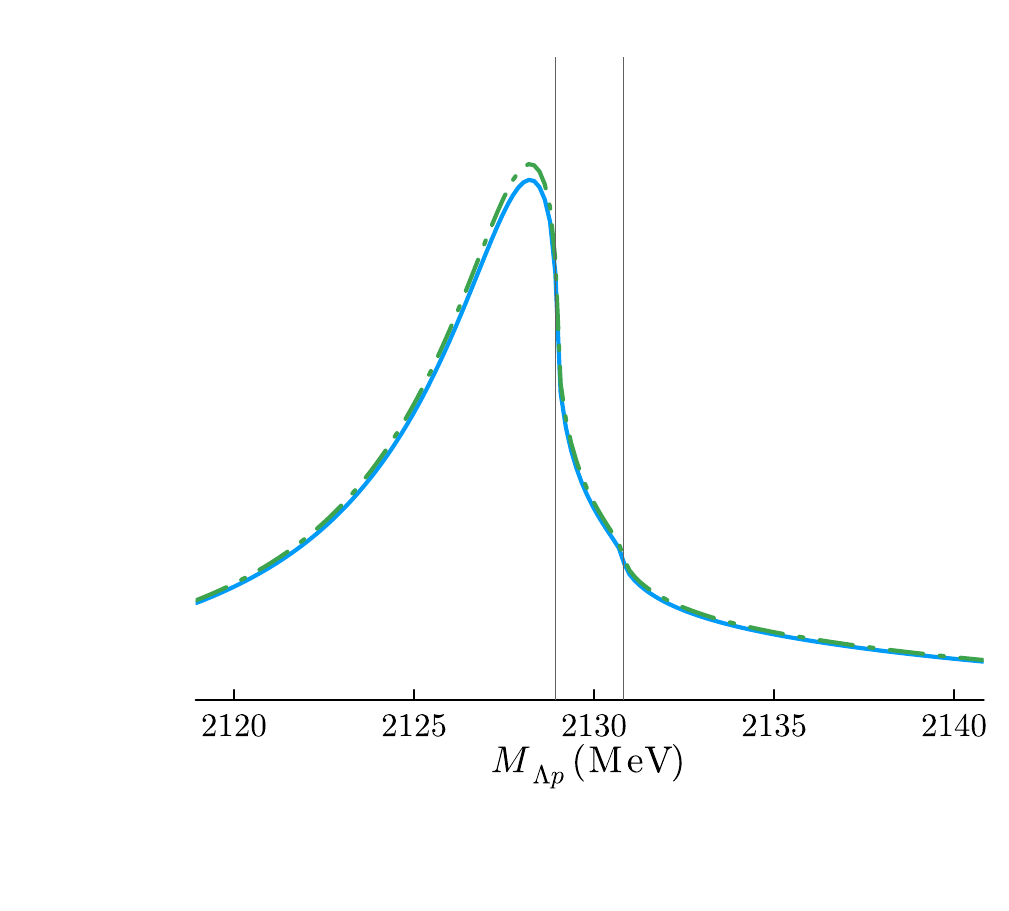}\par
  \vspace{1pt}{\footnotesize (b) Paris\par}
 \end{minipage}%
 \begin{minipage}[t]{0.31092\textwidth}\centering
  \includegraphics[width=\linewidth,trim=86.06 55 0 20,clip]{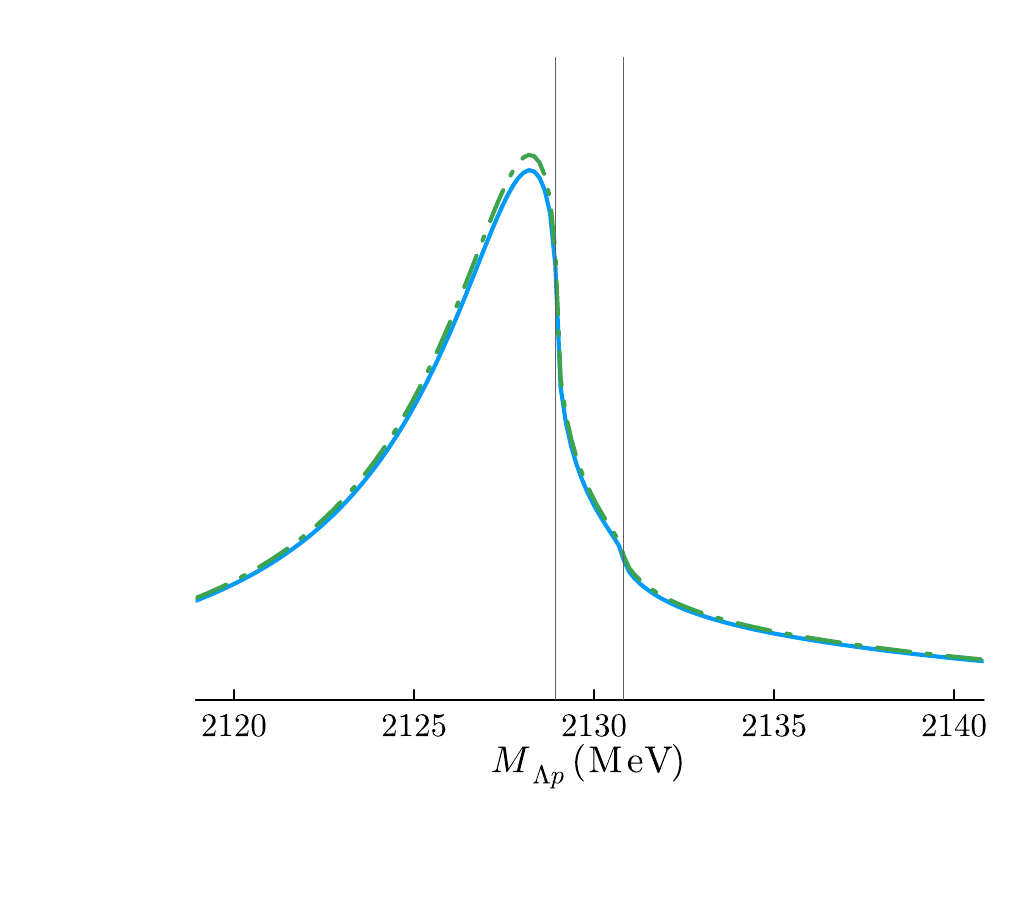}\par
  \vspace{1pt}{\footnotesize (c) AV18\par}
 \end{minipage}
 \caption{Dependence of the $\Lambda p$ invariant-mass spectrum on the deuteron wave function, using wave functions obtained from (a) the CD-Bonn potential, (b) the Paris potential, and (c) the AV18 potential.
 Only the $\Sigma$-exchange contributions are included.
 The solid curves include only the deuteron $s$-wave component, while the dash-dotted curves include both $s$- and $d$-wave components and their interference.
 The components are not independently renormalized.
 The calculations use the spin-triplet scattering lengths of Param.~3, $k_{\mathrm{lab}}=1.0$ GeV/$c$, and the angular selection $\cut$, with no proton-momentum cut.
 The spin-singlet $YN$ amplitude is omitted.
 The same vertical scale and line conventions apply to all three panels.
 The left and right vertical lines indicate the $\Sigma^+n$ and $\Sigma^0p$ thresholds, respectively.}\label{fig:deuteron}
\end{figure*}

We examine the dependence of the spectrum on the choice of deuteron wave function and on its $d$-wave component. Figure~\ref{fig:deuteron} shows the spectra obtained with wave functions from the CD-Bonn potential, the Paris potential~\cite{lacombe_1980}, and the Argonne $v_{18}$ (AV18) potential~\cite{wiringa_1995}. For this comparison, only the $\Sigma$-exchange contributions are included. The incident kaon momentum is 1.0 GeV/$c$, the spin-triplet scattering lengths are those of Param.~3, and the spin-singlet $YN$ amplitude is omitted. We apply the angular selection $\cut$ and impose no proton-momentum cut.

The solid curves are calculated with the deuteron $s$-wave component alone, while the dash-dotted curves also include the $d$-wave component. The original normalization is used for both components; in particular, the $s$-wave component is not renormalized when the $d$-wave component is omitted. The difference between the two curves includes the contribution of the $d$ wave and its interference with the $s$ wave. The reaction mechanisms and proton-momentum selection differ from those in Figs.~\ref{fig:spectrum} and \ref{fig:singlet}, so differences in the absolute magnitudes between these figures cannot be attributed solely to the deuteron wave function.

For all three wave functions, the $s$-wave component accounts for most of the spectrum. Including the deuteron $d$-wave component and its interference with the $s$ wave slightly increases the peak height, while leaving the peak position and the threshold line shape nearly unchanged. The $d$-wave component thus gives a small correction under the conditions considered here.

When both components are included, the spectra obtained using the Paris potential and the AV18 potential are similar and have lower peaks than the spectrum obtained using the CD-Bonn potential. The characteristic threshold structure remains visible for all three wave functions. Thus, in this comparison, the choice of deuteron wave function mainly affects the peak height rather than the qualitative line shape. Both the choice of wave function and the contribution of its $d$-wave component should be taken into account when using the strength of the enhancement to constrain the $\Sigma N$ interaction.

\section{Summary}\label{sec:sum}

We have examined the $\Lambda p$ invariant-mass spectrum near the two physical $\Sigma N$ thresholds in the $K^-d\to\pi^-\Lambda p$ reaction using a coupled-channel constant-$K$-matrix model. The line shape depends strongly on the complex spin-triplet $I=1/2$ $\Sigma N$ scattering length. The characteristic differences in the $\Sigma N\to\Lambda p$ conversion amplitude are closely related to the positions of nearby poles on unphysical Riemann sheets and are reflected in the reaction spectrum. The dependence on the $I=3/2$ scattering length is weaker overall. Its effect becomes noticeable near the lower $\Sigma^+n$ threshold owing to isospin breaking arising from the finite mass difference between the two $\Sigma N$ charge channels.

Param.~1 and Param.~3 produce broad enhancements below the $\Sigma^+n$ threshold, whereas Param.~2 gives a more localized cusp. These differences persist when the incident kaon momentum is increased from 1.0 to 1.4 GeV/$c$, with a reduction in the overall magnitude of the spectrum. The calculation at 760 MeV/$c$ qualitatively reproduces the suppression near the $\Lambda p$ threshold as the proton-momentum cut is increased, while the calculated spectra around the $\Sigma N$ thresholds remain nearly unchanged. The comparison with existing experimental data is qualitative, since the experimental 2 MeV bin width is not taken into account in the theoretical curves. It does not establish a preferred parameter set.

For the singlet scattering lengths and kinematic selections considered here, the spin-singlet $YN$ amplitude gives only a small correction near the main enhancement. A separate comparison of the $\Sigma$-exchange spectra using deuteron wave functions obtained from the CD-Bonn potential, the Paris potential, and the AV18 potential shows differences in the peak height, with similar threshold line shapes. For each wave function, including the $d$-wave component and its interference with the $s$ wave gives a small increase in the peak height, with little change in the peak position. The choice of deuteron wave function should therefore be taken into account in a quantitative analysis, while the $d$-wave component gives a small correction under the conditions considered here. The present results indicate that measurements of the line shape near the two $\Sigma N$ thresholds can provide useful constraints on the spin-triplet $I=1/2$ $\Sigma N$ interaction.

\section*{Acknowledgements}
We thank Kiyoshi Tanida and Yudai Ichikawa for valuable comments and discussions. S.Y. was supported in part by JST SPRING Grant No. JPMJSP2180. D.J. was supported in part by Grants-in-Aid for Scientific Research Grant Nos. JP22H04917, JP23K03427, and JP25K07315. Numerical computations were performed in part on the Yukawa-21 supercomputer at the Yukawa Institute for Theoretical Physics, Kyoto University. The authors used OpenAI’s GPT-6 Astra to assist with the analytic evaluation of the loop integrals and English-language editing. The resulting analytic expressions were checked by the authors against numerical integration, and the text was reviewed and revised.

\raggedbottom
\begin{appendix}\label{sec:app}
\section{INTEGRAL WITH THE DEUTERON WAVE FUNCTION}\label{sec:integral}
The loop integrals with the deuteron wave function are treated differently
for the hyperon- and meson-exchange processes. For hyperon exchange, the
energy component $q_0$ depends on the loop momentum, as in
Eq.~\eqref{equ:q0_hyp}, to reproduce the threshold position accurately.
After the angular integration, the remaining one-dimensional integral
is evaluated numerically. For meson exchange, the numerical calculation
also includes the bound-nucleon kinetic energy in $q_0$ and both
deuteron radial components. We first present the analytic loop functions
in the fixed-energy limit, followed by the angular reductions used with
the momentum-dependent propagators.

\subsection{Analytic expressions for the loop functions}\label{subsec:scalar_loops}
We define the loop functions for the deuteron $s$- and $d$-wave
components by
\begin{equation}
\Phi_n(Q,q_0;M) = \int\frac{d^3\vec{q}}{(2\pi)^3} \frac{P_n(\hat{q}\cdot\hat{Q})\ \varphi^{(0)}(|\vec{q}+\vec{Q}|)}{q_0^2 - |\vec{q}|^2 - M^2 + i\epsilon},
\end{equation}
and
\begin{equation}
    \Psi_n(Q,q_0;M)=\int\frac{d^3\vec{q}}{(2\pi)^3}\frac{P_n(\hat{q}\cdot\hat{Q})\ \varphi^{(2)}(|\vec{q}+\vec{Q}|)}{q_0^2-|\vec{q}|^2-M^2+i\epsilon}.
\end{equation}
Here $M$ is the mass of the exchanged particle and $P_n$ is a Legendre
polynomial. These definitions apply to both hyperon and meson exchange,
with $q_0$ allowed to depend on $\vec q$. For the analytic evaluation of
meson exchange, we take $q_0$ to be independent of the loop momentum
$\vec q$ and define
\begin{equation}
    \mu=\sqrt{M^2-q_0^2-\mathrm{i}\epsilon}.
\end{equation}
The square root is taken on the principal branch. The same prescription
is used in all closed forms below. Using the deuteron wave function,
$\Phi_n(Q,\mu)$ is parametrized as
\begin{equation}
\Phi_n(Q,\mu) = N\sum_{i} C_i I_n(Q,\mu;m_i),
\end{equation}
where $N$ is the common normalization factor and $C_i=C_i^{(0)}$ are the
$s$-wave coefficients defined in Sec.~\ref{subsec:deuteron}. The integral
$I_n(Q,\mu;m)$ is defined as
\begin{equation}
I_n(Q,\mu;m) = - \int\frac{d^3\vec{q}}{(2\pi)^3} \frac{P_n(\hat{q}\cdot\hat{Q})}{|\vec{q}|^2 + \mu^2}\frac{1}{(\vec{q}+\vec{Q})^2 +m^2}.  \label{equ:loop}
\end{equation}
\onecolumngrid
\vspace{\baselineskip}
By performing the Fourier transformation
\begin{equation}
\frac{1}{(\vec{q}+\vec{Q})^2 + m^2} = \int d^3\vec{r}\  \frac{e^{-m|\vec{r}|}}{4\pi |\vec{r}|}e^{-i(\vec{q}+\vec{Q})\cdot\vec{r}},
\end{equation}
and utilizing spherical harmonics, $I_n$ can be expressed as a double integral involving Bessel functions $J_n(z)$:
\begin{equation}
I_n(Q,\mu;m) = \frac{(-1)^{n+1}}{4\pi\sqrt{Q}}\int_0^\infty dq \Bigg[ \ \frac{q^{\frac{3}{2}}}{q^2+\mu^2}\int_0^\infty dr e^{-mr} J_{n+\frac{1}{2}}(Qr)J_{n+\frac{1}{2}}(qr)\Bigg].
\end{equation}
The integral over $r$ is the Laplace transform of a product of two Bessel functions and yields
\begin{equation}
\int_0^\infty dr \ e^{-mr} J_{n+\frac{1}{2}}(Qr) J_{n+\frac{1}{2}}(qr)=\frac{1}{\pi\sqrt{Qq}} Q_n\left(\frac{m^2+Q^2+q^2}{2Qq}\right),
\end{equation}
where $Q_n(z)$ is the Legendre function of the second kind.
Thus, the integral $I_n$ can be expressed as
\begin{equation}
I_n(Q,\mu;m) = \frac{(-1)^{n+1}}{4\pi^2Q}\int_0^\infty dq\frac{qQ_n\left(\frac{m^2+Q^2+q^2}{2Qq}\right)}{q^2+\mu^2}.
\end{equation}
The analytical expressions for $I_n$ with $n=0,\ldots,6$ are given below:
\begin{equation}
I_0(Q,\mu;m)= \frac{i}{8\pi Q}\text{Log}\left(\frac{m+\mu+Qi}{m+\mu-Qi}\right),
\end{equation}
\begin{equation}
    \begin{aligned}
        I_1(Q,\mu;m)={}&\frac{1}{4\pi^2 Q}\Bigg[1+\text{Log}\left(\frac{\mu}{\sqrt{m^2+Q^2}}\right)-\frac{m}{Q}\tan^{-1}\left(\frac{Q}{m}\right)+\frac{m^2+Q^2-\mu^2}{4Q}\mathcal B\left(Q,\mu;m\right)\Bigg],
    \end{aligned}
\end{equation}
\begin{equation}
    \begin{aligned}
        I_2(Q,\mu,m)={}&\frac{3}{32\pi\mu Q^2}\left(m^2+\mu m-\mu^2+Q^2\right) - \frac{3}{32\pi\mu^2Q^3}(m^2+Q^2)^2\arctan\left(\frac{Q}{m}\right)\\
                &-\frac{i}{64\pi\mu^2Q^3}\left\{3Q^4+2(3m^2-\mu^2)Q^2 + 3(m^2 - \mu^2)^2\right\}\text{Log}\left(\frac{m+\mu+Qi}{m+\mu-Qi}\right).
    \end{aligned}
\end{equation}
\begin{equation}
    \begin{aligned}
        I_3(Q,\mu;m)={}&\frac{1}{1152\pi^2Q^4\mu^2}\Bigg[
        4Q\Big\{45(m^2+Q^2)^2+\left(120(m^2+Q^2)-88Q^2\right)\mu^2-45\mu^4\Big\}\\
        &\qquad\qquad\qquad-12Q\Big\{15(m^2+Q^2)^2-\{30(m^2+Q^2)-16Q^2\}\mu^2+15\mu^4\Big\}\text{Log}\left(\frac{\mu}{\sqrt{m^2+Q^2}}\right)\\
        &\qquad\qquad\qquad-12m\Big\{15(m^2+Q^2)^2+\left(40(m^2+Q^2)-16Q^2\right)\mu^2-15\mu^4\Big\}\arctan\left(\frac{Q}{m}\right)\\
        &\qquad\qquad\qquad+9\left(\mu^2-(m^2+Q^2)\right)\Big\{5(m^2+Q^2-\mu^2)^2+12Q^2\mu^2\Big\}\mathcal B\left(Q,\mu;m\right)
        \Bigg].
    \end{aligned}
\end{equation}
\begin{equation}
    \begin{aligned}
        I_4(Q,\mu,m)={}&\frac{1}{3072\pi Q^5\mu^4}\Bigg[10Q\mu\Big\{-21(m^2+Q^2)^3+21m(m^2+Q^2)^2\mu +(m^2+Q^2) (63m^2+19Q^2)\mu^2\\
        &\qquad\qquad\qquad\qquad +3m(21m^2+11Q^2)\mu^3 -(63m^2+19Q^2)\mu^4-21m\mu^5+21\mu^6\Big\}{}\\
        &\qquad\qquad\qquad+30(m^2+Q^2)^2\left\{7(m^2+Q^2)^2-4(7m^2+Q^2)\mu^2\right\}\arctan\left(\frac{Q}{m}\right)\\
        &\qquad\qquad\qquad+3\mathrm{i}\Big\{35(m^2+Q^2)^4-20(m^2+Q^2)^2(7m^2+Q^2)\mu^2\\
        &\qquad\qquad\qquad\qquad+6(35m^4+30m^2Q^2+3Q^4)\mu^4-20(7m^2+Q^2)\mu^6+35\mu^8\Big\}\text{Log}\left(\frac{m+\mu+\mathrm{i}Q}{m+\mu-\mathrm{i}Q}\right)\Bigg].
    \end{aligned}
\end{equation}
\begingroup
\begin{equation}
    \begin{aligned}
        I_5(Q,\mu;m)={}&\frac{1}{460800\pi^2Q^6\mu^4}\Bigg[
        4Q\Big\{14175\left[\mu^8-(m^2+Q^2)^4\right]\\
        &\qquad+9450(7m^2+Q^2)\left[(m^2+Q^2)^2\mu^2-\mu^6\right]
        +128(945m^4+735m^2Q^2+64Q^4)\mu^4\Big\}\\
        &\quad+60Q\Big\{945(m^2+Q^2-\mu^2)^4
        +2940Q^2\mu^2(m^2+Q^2-\mu^2)^2+1024Q^4\mu^4\Big\}
        \text{Log}\left(\frac{\mu}{\sqrt{m^2+Q^2}}\right)\\
        &\quad+60m\Big\{945\left[(m^2+Q^2)^4-\mu^8\right]
        -1470(3m^2+Q^2)\left[(m^2+Q^2)^2\mu^2-\mu^6\right]\\
        &\qquad-128(63m^4+70m^2Q^2+15Q^4)\mu^4\Big\}
        \arctan\left(\frac{Q}{m}\right)\\
        &\quad+225(m^2+Q^2-\mu^2)\Big\{63(m^2+Q^2-\mu^2)^4
        +280Q^2\mu^2(m^2+Q^2-\mu^2)^2+240Q^4\mu^4\Big\}
        \mathcal B\left(Q,\mu;m\right)\Bigg].
    \end{aligned}
\end{equation}
\begin{equation}
    \begin{aligned}
        I_6(Q,\mu;m)={}&\frac{1}{40960\pi Q^7\mu^6}\Bigg[
        14Q\mu\Big\{165(m^2+Q^2)^5-165m(m^2+Q^2)^4\mu\\
        &\qquad-5(m^2+Q^2)^3(165m^2+29Q^2)\mu^2
        +15m(m^2+Q^2)^2(55m^2+17Q^2)\mu^3\\
        &\qquad+6(m^2+Q^2)(275m^4+210m^2Q^2+23Q^4)\mu^4
        +10m(165m^4+170m^2Q^2+33Q^4)\mu^5\\
        &\qquad-6(275m^4+210m^2Q^2+23Q^4)\mu^6
        -15m(55m^2+17Q^2)\mu^7\\
        &\qquad+5(165m^2+29Q^2)\mu^8+165m\mu^9-165\mu^{10}\Big\}\\
        &\quad-210(m^2+Q^2)^2\Big\{11(m^2+Q^2)^4
        -6(m^2+Q^2)^2(11m^2+Q^2)\mu^2+5(33m^4+18m^2Q^2+Q^4)\mu^4\Big\}
        \arctan\left(\frac{Q}{m}\right)\\
        &\quad-5\mathrm{i}\Big\{231(m^2+Q^2-\mu^2)^6
        +1260Q^2\mu^2(m^2+Q^2-\mu^2)^4\\
        &\qquad+1680Q^4\mu^4(m^2+Q^2-\mu^2)^2+320Q^6\mu^6\Big\}
        \text{Log}\left(\frac{m+\mu+\mathrm{i}Q}{m+\mu-\mathrm{i}Q}\right)\Bigg],
    \end{aligned}
\end{equation}
\endgroup
where the auxiliary function $\mathcal B(Q,\mu;m)$ collects the logarithmic and dilogarithmic $\operatorname{Li}_2$ terms common to odd $I_n$ and is written in terms of the variables $z_\pm$ as follows:
\begingroup
\begin{equation}\label{equ:auxiliary_B}
    \begin{aligned}
        \mathcal B(Q,\mu;m)={}&\frac{\mathrm{i}}{\mu}\Bigg[
        \operatorname{Li}_2(1-z_+)-\operatorname{Li}_2(1-z_-)+\operatorname{Li}_2(-z_+)-\operatorname{Li}_2(-z_-)+\operatorname{Log}z_+\operatorname{Log}(1+z_+)-\operatorname{Log}z_-\operatorname{Log}(1+z_-)\Bigg],\\
        z_\pm={}&\frac{m\pm\mathrm{i}Q}{\mu}.
    \end{aligned}
\end{equation}
\endgroup
The argument of $\mathcal B$ is evaluated with the same physical
boundary value of $\mu$ defined above. The infinitesimal prescription
is used when evaluating the logarithms and dilogarithms across their
branch cuts.

\subsection{Angular integrals and partial-wave contractions}\label{subsec:angular_loops}
The angular reductions below apply to both fixed and momentum-dependent
$q_0$, with $\Phi_n$ defined by the corresponding integrals above.
Products of angular factors in the meson-exchange amplitudes
lead to tensor loop integrals. For the deuteron $s$-wave contribution,
$\varphi$ in the following definition denotes $\varphi^{(0)}$.
Because the scalar kernel depends on directions only through
$\hat q\cdot\hat Q$, each tensor can be decomposed into Kronecker deltas
and components of $\hat Q$, with coefficients given by the moments
$\Phi_n$.
\begin{equation}
    \begin{aligned}
        \mathcal{I}_{i_1\cdots i_n}&=\int\frac{d^3\vec{q}}{(2\pi)^3}\frac{\hat{q}_{i_1}\cdots\hat{q}_{i_n}}{q_0^2-|\vec q|^2-M^2+i\epsilon}\varphi(|\vec{q}+\vec{Q}|)\\
        &=\sum_{k=0}^{\lfloor n/2\rfloor}\frac{n!(2n-4k+1)!!}{(2k)!!(n-2k)!(2n-2k+1)!!}\delta_{(i_1i_2}\cdots\delta_{i_{2k-1}i_{2k}}\int\frac{d^3\vec{q}}{(2\pi)^3}\frac{\varphi(|\vec{q}+\vec{Q}|)}{q_0^2-|\vec q|^2-M^2+i\epsilon}\hat{q}_{\langle i_{2k+1}\cdots i_n\rangle)}\\
        &=\sum_{k=0}^{\lfloor n/2\rfloor}\frac{n!(2n-4k+1)!!}{(2k)!!(n-2k)!(2n-2k+1)!!}\delta_{(i_1i_2}\cdots\delta_{i_{2k-1}i_{2k}}\hat{Q}_{\langle i_{2k+1}\cdots i_n\rangle)}\Phi_{n-2k}(Q)\label{equ:tensor_loop}
    \end{aligned}
\end{equation}
Parentheses on indices denote normalized symmetrization, and
angle brackets denote the symmetric trace-free part. Repeated Cartesian
indices are summed. The explicit forms through rank six are listed
below; the ellipses include all distinct permutations of the displayed
index structures.
\begingroup
\allowdisplaybreaks[4]
\begin{align}
    \mathcal{I}_i=&\ \hat{Q}_i\Phi_1\notag\\
    \mathcal{I}_{ij}=&\ \delta_{ij}\frac{1}{3}(\Phi_0-\Phi_2) + \hat{Q}_i\hat{Q}_j\Phi_2\notag\\
    \mathcal{I}_{ijk}=&\left(\delta_{ij}\hat{Q}_k + \delta_{jk}\hat{Q}_i + \delta_{ki}\hat{Q}_j\right)\frac{1}{5}(\Phi_1-\Phi_3) + \hat{Q}_i\hat{Q}_j\hat{Q}_k\Phi_3\notag\\
    \mathcal{I}_{ijkl}=&\ \Big(\delta_{ij}\delta_{kl} + \delta_{ik}\delta_{jl} + \delta_{il}\delta_{jk}\Big)\frac{1}{105}(7\Phi_0 - 10\Phi_2 + 3\Phi_4)+\left(\delta_{ij}\hat{Q}_k\hat{Q}_l + \cdots\right)\frac{1}{7}(\Phi_2-\Phi_4)+\hat{Q}_i\hat{Q}_j\hat{Q}_k\hat{Q}_l\Phi_4\notag\\
    \mathcal{I}_{ijklm}=&\ \Big(\delta_{ij}\delta_{kl}\hat{Q}_m + \cdots\Big)\frac{1}{315}(9\Phi_1 - 14\Phi_3 + 5\Phi_5)+\left(\delta_{ij}\hat{Q}_k\hat{Q}_l\hat{Q}_m + \cdots\right)\frac{1}{9}(\Phi_3-\Phi_5)+\hat{Q}_i\hat{Q}_j\hat{Q}_k\hat{Q}_l\hat{Q}_m\Phi_5\notag\\
    \mathcal{I}_{ijklmn}=&\ \Big(\delta_{ij}\delta_{kl}\delta_{mn} + \cdots \Big)\frac{1}{3465}(33\Phi_0 - 55\Phi_2 + 27\Phi_4 - 5\Phi_6)+\left(\delta_{ij}\delta_{kl}\hat{Q}_m\hat{Q}_n + \cdots\right)\frac{1}{693}(11\Phi_2 - 18\Phi_4 + 7\Phi_6)\notag\\
    &\ +\left(\delta_{ij}\hat{Q}_k\hat{Q}_l\hat{Q}_m\hat{Q}_n + \cdots\right)\frac{1}{11}(\Phi_4-\Phi_6)+\hat{Q}_i\hat{Q}_j\hat{Q}_k\hat{Q}_l\hat{Q}_m\hat{Q}_n\Phi_6
\end{align}
\endgroup
The angular factors at the two meson--baryon vertices can be
contracted with these tensors. Let $\hat k_1$ and $\hat k_2$ denote
the initial and final meson directions, respectively. We introduce
the three scalar products
\begin{equation}
    u=\hat k_2\cdot\hat Q,\qquad
    v=\hat Q\cdot\hat k_1,\qquad
    w=\hat k_2\cdot\hat k_1,\label{equ:angular_invariants}
\end{equation}
and define the corresponding partial-wave loop function by
\begin{equation}
    \mathcal J_{\ell_2\ell_1}(u,v,w)=\int\frac{d^3\vec q}{(2\pi)^3}\frac{\varphi^{(0)}(|\vec q+\vec Q|)}{q_0^2-|\vec q|^2-M^2+i\epsilon}P_{\ell_2}(\hat k_2\cdot\hat q)P_{\ell_1}(\hat q\cdot\hat k_1).\label{equ:j_loop}
\end{equation}
The remaining kinematic dependence is suppressed in the
notation for $\mathcal J_{\ell_2\ell_1}$ and $\Phi_n$. Expanding the
Legendre polynomials into products of Cartesian components reduces
$\mathcal J$ to contractions of $\mathcal I$; for example,
$\mathcal J_{11}=\hat k_{2i}\hat k_{1j}\mathcal I_{ij}$. With
$\ell_1,\ell_2\leq3$, these contractions require tensor ranks and
scalar moments only up to six. The explicit expressions are
\begin{equation}
    \begin{aligned}
        \mathcal J_{00}&=\Phi_0,\label{equ:j_explicit}\\
        \qquad\mathcal J_{01}&=v\Phi_1,\qquad\mathcal J_{10}=u\Phi_1,\\
        \mathcal J_{02}&=\frac{1}{2}\left(3v^2-1\right)\Phi_2,
        \qquad\mathcal J_{20}=\frac{1}{2}\left(3u^2-1\right)\Phi_2,
        \qquad\mathcal J_{11}=\frac{1}{3}w\left(\Phi_0-\Phi_2\right)+uv\Phi_2,\\
        \mathcal J_{03}&=\frac{1}{2}v\left(5v^2-3\right)\Phi_3,
        \qquad\mathcal J_{30}=\frac{1}{2}u\left(5u^2-3\right)\Phi_3,\\
        \mathcal J_{12}&=\frac{1}{5}\left(3vw-u\right)\Phi_1+\frac{3}{10}\left(5uv^2-u-2vw\right)\Phi_3,
        \qquad\mathcal J_{21}=\frac{1}{5}\left(3uw-v\right)\Phi_1+\frac{3}{10}\left(5u^2v-2uw-v\right)\Phi_3,\\
        \mathcal J_{13}&=-\frac{3}{14}\left(2uv-5v^2w+w\right)\Phi_2+\frac{1}{14}\left(35uv^3-15uv-15v^2w+3w\right)\Phi_4,\\
        \mathcal J_{31}&=-\frac{3}{14}\left(2uv-5u^2w+w\right)\Phi_2+\frac{1}{14}\left(35u^3v-15u^2w-15uv+3w\right)\Phi_4,\\
        \mathcal J_{22}&=\frac{1}{10}\left(3w^2-1\right)\Phi_0-
        \frac{1}{7}\left(3u^2-9uvw+3v^2+3w^2-2\right)\Phi_2\\
        &\quad+\frac{9}{140}\left(35u^2v^2-5u^2-20uvw-5v^2+2w^2+1\right)\Phi_4,\\
        \mathcal J_{23}&=-\frac{9}{70}\left(2uw-5vw^2+v\right)\Phi_1-\frac{1}{30}\left(30u^2v-75uv^2w+3uw+25v^3+30vw^2-21v\right)\Phi_3\\
        &\quad+\frac{5}{84}\left(63u^2v^3-21u^2v-42uv^2w+6uw-7v^3+6vw^2+3v\right)\Phi_5,\\
        \mathcal J_{32}&=-\frac{9}{70}\left(-5uw^2+u+2vw\right)\Phi_1-\frac{1}{30}\left(25u^3-75u^2vw+30uv^2+30uw^2-21u+3vw\right)\Phi_3\\
        &\quad+\frac{5}{84}\left(63u^3v^2-7u^3-42u^2vw-21uv^2+6uw^2+3u+6vw\right)\Phi_5,\\
        \mathcal J_{33}&=\frac{1}{14}w\left(5w^2-3\right)\Phi_0-\frac{1}{42}\left(30u^2w-75uvw^2+3uv+30v^2w+25w^3-21w\right)\Phi_2\\
        &\quad+\frac{9}{308}\left(-70u^3v+175u^2v^2w+5u^2w-70uv^3-100uvw^2+60uv+5v^2w+10w^3-7w\right)\Phi_4\\
        &\quad+\frac{25}{924}\left(231u^3v^3-63u^3v-189u^2v^2w+21u^2w-63uv^3+42uvw^2+21uv+21v^2w-2w^3-3w\right)\Phi_6.
    \end{aligned}
\end{equation}
\section{DETAILS OF THE AMPLITUDE}\label{sec:amp_details}
We decompose the reaction amplitudes in baryon spin space as
\begin{equation}
    \mathcal{T}_a = AS_a^\dag + iB_iS_a^\dag(\sigma_i)^T + iC_i\sigma_iS_a^\dag + D_{ij}\sigma_iS_a^\dag(\sigma_j)^T,\\
\end{equation}
where $S_a^\dag$ is defined in Sec.~\ref{subsec:deuteron}, $\sigma_i$
are Pauli matrices, and $T$ denotes matrix transposition.
For contributions $\alpha$ and $\beta$, the spin-averaged product is
given below, with a sum over the three deuteron spin states $a$ and
repeated Cartesian indices; an asterisk denotes complex conjugation.
\begin{equation}
\begin{aligned}
    \frac{1}{3}\sum_{a=1}^{3}\operatorname{tr}[\mathcal{T}^{(\alpha)}_a\mathcal{T}^{(\beta)\dag}_a]
    =&\ A^{(\alpha)}A^{*(\beta)} + B^{(\alpha)}_iB^{*(\beta)}_i + C^{(\alpha)}_iC^{*(\beta)}_i + D^{(\alpha)}_{ij}D^{*(\beta)}_{ij}+\frac{1}{3}(D^{(\alpha)}_{ij}D^{*(\beta)}_{ji}-D^{(\alpha)}_{ii}D^{*(\beta)}_{jj})\\
    &+\frac{1}{3}(A^{(\alpha)}D^{*(\beta)}_{ii}+D^{(\alpha)}_{ii}A^{*(\beta)})
    +\frac{1}{3}(B^{(\alpha)}_iC^{*(\beta)}_i+C^{(\alpha)}_iB^{*(\beta)}_i)\\
    &+\frac{1}{3}\varepsilon_{ijk}(-B_i^{(\alpha)}D^{*(\beta)}_{jk}-D^{(\alpha)}_{ij}B^{*(\beta)}_k)
    +\frac{1}{3}\varepsilon_{ijk}(C_i^{(\alpha)}D^{*(\beta)}_{jk}+D^{(\alpha)}_{ij}C^{*(\beta)}_k).
\end{aligned}
\end{equation}
\subsection{Impulse process}
Using the deuteron spin matrices in Sec.~\ref{subsec:deuteron},
the impulse amplitude in Eq.~\eqref{equ:imp} is written as
\begin{equation*}
    \mathcal{T}^{(I)}_a=(F-iG\vec{n}\cdot\vec\sigma) S_a^\dag \varphi^{(0)}(|\vec{p}_N|) + (F-iG\vec{n}\cdot\vec\sigma)\left[\left(\frac{3}{2\sqrt2}
    (\hat{p}_N)_i(\hat{p}_N)_j-\frac{1}{2\sqrt{2}}\delta_{ij}\right)\sigma_iS^\dag_a(\sigma_j)^T\right] \varphi^{(2)}(|\vec{p}_N|)
\end{equation*}
The corresponding coefficients are
\begin{equation}
    \begin{aligned}
        A^{(I)}&=F\varphi^{(0)}(|\vec{p}_N|),\\
        B^{(I)}_i&=\frac{G}{2\sqrt2}\left[-3(\vec{n}\cdot\hat{p}_N)(\hat{p}_N)_i + n_i\right]\varphi^{(2)}(|\vec{p}_N|),\\
        C^{(I)}_i&=-Gn_i\varphi^{(0)}(|\vec{p}_N|),\\
        D^{(I)}_{ij}&=\frac{1}{2\sqrt2}\left[F(3(\hat{p}_N)_i(\hat{p}_N)_j-\delta_{ij})+G(3(\vec{n}\times\hat{p}_N)_i(\hat{p}_N)_j + \varepsilon_{ijk}n_k)\right]\varphi^{(2)}(|\vec{p}_N|),\\
    \end{aligned}
\end{equation}
\subsection{Hyperon exchange process}
The hyperon-exchange amplitude is
\begin{align*}
    \mathcal{T}^{(Y)}_a=&
    \int\frac{d^3\vec{q}}{(2\pi)^3}\frac{2M_Y}{q^2-M_Y^2+i\varepsilon}\varphi^{(0)}(|\vec{q}+\vec{Q}|)\left[T^1_{YN}(F-iG\vec{n}\cdot\vec\sigma)S^{\dag}_a+ T^\sigma_{YN} \sum_b \sigma_b (F - iG\vec{n}\cdot\vec\sigma)S^\dag_a (\sigma_b)^T\right]\\
    &+\int\frac{d^3\vec{q}}{(2\pi)^3}\frac{2M_Y}{q^2-M_Y^2+i\varepsilon}\varphi^{(2)}(|\vec{q}+\vec{Q}|)
    \Bigg[T_{YN}^1(F-iG\vec{n}\cdot\vec\sigma)\left\{\frac{3}{2\sqrt2}(\widehat{q+Q})_i(\widehat{q+Q})_j-\frac{1}{2\sqrt{2}}\delta_{ij}\right\}\sigma_iS^\dag_a(\sigma_j)^T\\
    &\qquad +T_{YN}^\sigma \sum_b\sigma_b(F-iG\vec{n}\cdot\vec\sigma)\left\{\frac{3}{2\sqrt2}(\widehat{q+Q})_i(\widehat{q+Q})_j-\frac{1}{2\sqrt{2}}\delta_{ij}\right\}\sigma_iS^\dag_a(\sigma_j)^T(\sigma_b)^T\Bigg]\\
\end{align*}
The loop functions are evaluated as in Appendix~\ref{sec:integral},
with $M=M_Y$ and $q_0$ given by Eq.~\eqref{equ:q0_hyp}.
The $d$-wave angular factor is $P_2(\hat p\cdot\hat Q)$,
with the deuteron momentum $\vec p=\vec q+\vec Q$.
$F$, $G$, and
$\vec n=\hat p_{\pi^-}\times\hat p_{K^-}$ are evaluated at the external
kinematics and held fixed during the loop integration. The resulting
coefficients are
\begin{equation}
    \begin{aligned}
        A^{(Y)}={}&2M_YT^t_{YN}F\Phi_0^{(Y)},\\
        B^{(Y)}_i=&-M_Y\left(T^t_{YN}-T^s_{YN}\right)Gn_i\Phi_0^{(Y)}+\frac{M_Y}{2\sqrt{2}}\left(T^t_{YN}+T^s_{YN}\right)G\Psi_2^{(Y)}\left[n_i-3(\vec n\cdot\hat Q)\hat Q_i\right],\\
        C^{(Y)}_i=&-M_Y\left(T^t_{YN}+T^s_{YN}\right)Gn_i\Phi_0^{(Y)}+\frac{M_Y}{2\sqrt{2}}\left(T^t_{YN}-T^s_{YN}\right)G\Psi_2^{(Y)}\left[n_i-3(\vec n\cdot\hat Q)\hat Q_i\right],\\
        D^{(Y)}_{ij}={}&\frac{M_Y}{\sqrt{2}}T^t_{YN}F\Psi_2^{(Y)}\left(3\hat Q_i\hat Q_j-\delta_{ij}\right)\\
        &+\frac{M_Y}{2\sqrt{2}}
        G\Psi_2^{(Y)}
        \Bigg[6T^t_{YN}(\vec n\times\hat Q)_i\hat Q_j
        +(3T^t_{YN}-T^s_{YN})\varepsilon_{ijk}n_k
        -3\left(T^t_{YN}-T^s_{YN}\right)\varepsilon_{ijk}\hat Q_k(\vec n\cdot\hat Q)\Bigg].
    \end{aligned}
\end{equation}
\subsection{Meson exchange process}
The subscripts 1 and 2 label the first and second scatterings, with
$\vec n_1=\hat q\times\hat k_1$ and
$\vec n_2=\hat k_2\times\hat q$.
We use $q^2=q_0^2-|\vec q|^2$, with $q_0$ given by
Eqs.~\eqref{equ:q0_k} and \eqref{equ:q0_pi}:
\begin{equation*}
    \begin{aligned}
        \mathcal{T}^{(\pi,K^-)}_a=&\int\frac{d^3\vec{q}}{(2\pi)^3}\frac{\varphi^{(0)}(|\vec q+\vec Q|)}{q^2-M_{(\pi,K^-)}^2+i\varepsilon} {}(F_2-iG_2\vec{n}_2\cdot\vec\sigma)S_a^\dag(F_1-iG_1\vec{n}_1\cdot\vec\sigma)^T\\
        \quad&+\int\frac{d^3\vec{q}}{(2\pi)^3}\frac{\varphi^{(2)}(|\vec q+\vec Q|)}{q^2-M_{(\pi,K^-)}^2+i\varepsilon} {}(F_2-iG_2\vec{n}_2\cdot\vec\sigma)\left(\frac{3}{2\sqrt2}(\widehat{q+Q})_i(\widehat{q+Q})_j-\frac{1}{2\sqrt{2}}\delta_{ij}\right)\sigma_iS_a^\dag(\sigma_j)^T(F_1-iG_1\vec{n}_1\cdot\vec\sigma)^T
    \end{aligned}
\end{equation*}
The coefficients below describe the deuteron $s$-wave term, using
$\mathcal J_{\ell_2\ell_1}$ from Eq.~\eqref{equ:j_loop}.
Here $F_r^{(\ell)}$ and $G_r^{(\ell)}$ are the coefficients of $P_\ell$
and $P'_\ell$, respectively. Derivatives with respect to $u$, $v$, and
$w$ act on the angular factors with the loop kernel fixed.
The $d$-wave term is evaluated numerically using the full spin--angular
tensor in the second integral.
\begin{equation}
\begin{aligned}
    A^{(\pi,K^-)}=&\sum_{\ell_2=0}^{3}\sum_{\ell_1=0}^{3}F_2^{(\ell_2)}F_1^{(\ell_1)}\mathcal J_{\ell_2\ell_1},\\
    B^{(\pi,K^-)}_i=&-\sum_{\ell_2=0}^{3}\sum_{\ell_1=1}^{3}F_2^{(\ell_2)}G_1^{(\ell_1)}\left[\frac{\partial\mathcal J_{\ell_2\ell_1}}{\partial v}(\hat Q\times\hat k_1)_i+\frac{\partial\mathcal J_{\ell_2\ell_1}}{\partial w}(\hat k_2\times\hat k_1)_i\right],\\
    C^{(\pi,K^-)}_i=&-\sum_{\ell_2=1}^{3}\sum_{\ell_1=0}^{3}G_2^{(\ell_2)}F_1^{(\ell_1)}\left[\frac{\partial\mathcal J_{\ell_2\ell_1}}{\partial u}(\hat k_2\times\hat Q)_i+\frac{\partial\mathcal J_{\ell_2\ell_1}}{\partial w}(\hat k_2\times\hat k_1)_i\right],\\
    D^{(\pi,K^-)}_{ij}=&-\sum_{\ell_2=1}^{3}\sum_{\ell_1=1}^{3}G_2^{(\ell_2)}G_1^{(\ell_1)}\Bigg[
        \frac{\partial^2\mathcal J_{\ell_2\ell_1}}{\partial u\,\partial v}
        (\hat k_2\times\hat Q)_i
        (\hat Q\times\hat k_1)_j+
        \frac{\partial^2\mathcal J_{\ell_2\ell_1}}{\partial u\,\partial w}
        (\hat k_2\times\hat Q)_i
        (\hat k_2\times\hat k_1)_j
        \\&+
        \frac{\partial^2\mathcal J_{\ell_2\ell_1}}{\partial v\,\partial w}
        (\hat k_2\times\hat k_1)_i
        (\hat Q\times\hat k_1)_j+
        \frac{\partial^2\mathcal J_{\ell_2\ell_1}}{\partial w^2}
        (\hat k_2\times\hat k_1)_i
        (\hat k_2\times\hat k_1)_j+
        \frac{\partial\mathcal J_{\ell_2\ell_1}}{\partial w}
        \left(\hat k_{1i}\hat k_{2j}-w\delta_{ij}\right)\Bigg].
\end{aligned}
\end{equation}
\end{appendix}
\flushbottom
\twocolumngrid

\bibliographystyle{apsrev4-2}
\bibliography{Reference}

@article{acharya_2019,
  title = {$p\ensuremath{-}p, p\ensuremath{-}\mathrm{\ensuremath{\Lambda}}$, and $\mathrm{\ensuremath{\Lambda}}\ensuremath{-}\mathrm{\ensuremath{\Lambda}}$ correlations studied via femtoscopy in $pp$ reactions at $\sqrt{s}=7\phantom{\rule{0.16em}{0ex}}\mathrm{TeV}$},
  author = {Acharya, S. and others},
  collaboration = {ALICE Collaboration},
  journal = {Phys. Rev. C},
  volume = {99},
  issue = {2},
  pages = {024001},
  numpages = {21},
  year = {2019},
  month = {Feb},
  publisher = {American Physical Society},
  doi = {10.1103/PhysRevC.99.024001},
  url = {https://link.aps.org/doi/10.1103/PhysRevC.99.024001}
}

@article{acharya_2020,
  title = {Investigation of the p–Σ0 interaction via femtoscopy in pp collisions},
  journal = {Physics Letters B},
  volume = {805},
  pages = {135419},
  year = {2020},
  issn = {0370-2693},
  doi = {10.1016/j.physletb.2020.135419},
  url = {https://www.sciencedirect.com/science/article/pii/S0370269320302239},
  author = {S. Acharya and others}
}

@article{adamczewski-musch_2016,
  title = {$\mathrm{\ensuremath{\Lambda}}p$ interaction studied via femtoscopy in $p\phantom{\rule{0.28em}{0ex}}+\phantom{\rule{0.28em}{0ex}}\mathrm{Nb}$ reactions at $\sqrt{{s}_{\mathit{NN}}}=3.18$ GeV},
  author = {Adamczewski-Musch, J. and others},
  collaboration = {HADES Collaboration},
  journal = {Phys. Rev. C},
  volume = {94},
  issue = {2},
  pages = {025201},
  numpages = {12},
  year = {2016},
  month = {Aug},
  publisher = {American Physical Society},
  doi = {10.1103/PhysRevC.94.025201},
  url = {https://link.aps.org/doi/10.1103/PhysRevC.94.025201}
}

@article{aikawa_2023,
  title = {Pole Position of {{$\Lambda$}}(1405) Measured in d({{K}}-,n){{$\pi\Sigma$}} Reactions},
  author = {Aikawa, S. and others},
  collaboration = {J-PARC E31 Collaboration},
  year = {2023},
  month = feb,
  journal = {Physics Letters B},
  volume = {837},
  pages = {137637},
  issn = {0370-2693},
  urldate = {2023-05-17},
  langid = {english},
  doi = {10.1016/j.physletb.2022.137637},
  url = {https://doi.org/10.1016/j.physletb.2022.137637},
}

@article{akaishi_2002,
  title = {Ab initio Approach to $s$-Shell Hypernuclei $_{\mathrm{\ensuremath{\Lambda}}}^{\mathrm{3}}\mathrm{H}$, $_{\mathrm{\ensuremath{\Lambda}}}^{\mathrm{4}}\mathrm{H}$, $_{\mathrm{\ensuremath{\Lambda}}}^{\mathrm{4}}\mathrm{H}\mathrm{e}$, and $_{\mathrm{\ensuremath{\Lambda}}}^{\mathrm{5}}\mathrm{H}\mathrm{e}$ with a $\ensuremath{\Lambda}N\mathrm{\text{\ensuremath{-}}}\ensuremath{\Sigma}N$ Interaction},
  author = {Nemura, H. and Akaishi, Y. and Suzuki, Y.},
  journal = {Phys. Rev. Lett.},
  volume = {89},
  issue = {14},
  pages = {142504},
  numpages = {4},
  year = {2002},
  month = {Sep},
  publisher = {American Physical Society},
  doi = {10.1103/PhysRevLett.89.142504},
  url = {https://link.aps.org/doi/10.1103/PhysRevLett.89.142504}
}

@article{alexander_1968,
  title = {Study of the $\ensuremath{\Lambda}\ensuremath{-}N$ System in Low-Energy $\ensuremath{\Lambda}\ensuremath{-}p$ Elastic Scattering},
  author = {Alexander, G. and Karshon, U. and Shapira, A. and Yekutieli, G. and Engelmann, R. and Filthuth, H. and Lughofer, W.},
  journal = {Phys. Rev.},
  volume = {173},
  issue = {5},
  pages = {1452--1460},
  numpages = {0},
  year = {1968},
  month = {Sep},
  publisher = {American Physical Society},
  doi = {10.1103/PhysRev.173.1452},
  url = {https://link.aps.org/doi/10.1103/PhysRev.173.1452}
}

@article{alexander_1969,
  title = {Study of the {{Enhancement}} in the \$\textbackslash ensuremath\{\textbackslash\vphantom\}{{Lambda}}\vphantom\{\}p\$ {{Invariant Mass Near}} the \$\textbackslash ensuremath\{\textbackslash\vphantom\}{{Sigma}}\vphantom\{\}\textbackslash ensuremath\{\textbackslash\vphantom\}{{Lambda}}\vphantom\{\}\$ {{Threshold}} in \$\{\vphantom\}{{K}}\vphantom\{\}\^\{\textbackslash ensuremath\{-\}\}d\textbackslash ensuremath\{\textbackslash rightarrow\}\textbackslash ensuremath\{\textbackslash\vphantom\}{{Lambda}}\vphantom\{\}p\{\textbackslash ensuremath\{\textbackslash pi\}\}\^\{\textbackslash ensuremath\{-\}\}\$ {{Interactions Around}} 1.0 {{GeV}}/c},
  author = {Alexander, Gideon and Hall, Bronwyn H. and Jew, Nathan and Kalmus, George and Kernan, Anne},
  year = {1969},
  month = mar,
  journal = {Phys. Rev. Lett.},
  volume = {22},
  number = {10},
  pages = {483--486},
  publisher = {{American Physical Society}},
  urldate = {2023-05-15},
  doi = {10.1103/PhysRevLett.22.483},
  url = {https://doi.org/10.1103/PhysRevLett.22.483},
}

@article{beane_2007,
title = {Hyperon–nucleon scattering from fully-dynamical lattice QCD},
journal = {Nuclear Physics A},
collaboration = {NPLQCD Collaboration},
volume = {794},
number = {1},
pages = {62-72},
year = {2007},
issn = {0375-9474},
doi = {10.1016/j.nuclphysa.2007.07.006},
url = {https://www.sciencedirect.com/science/article/pii/S0375947407006434},
author = {Silas R. Beane and Paulo F. Bedaque and Thomas C. Luu and Kostas Orginos and Elisabetta Pallante and Assumpta Parreño and Martin J. Savage}
}

@article{braun_1977,
  title = {On the {{$\Lambda$p}} Enhancement near {{$\Sigma$N}} Threshold},
  author = {Braun, O. and Grimm, H. J. and Hepp, V. and Stroebele, H. and Thoel, C. and Thouw, T. J. and Gandini, F. and Kiesling, C. and Plane, D. E. and Wittek, W.},
  year = {1977},
  month = jun,
  journal = {Nuclear Physics B},
  volume = {124},
  number = {1},
  pages = {45--60},
  issn = {0550-3213},
  urldate = {2023-06-14},
  langid = {english},
  doi = {10.1016/0550-3213(77)90275-9},
  url = {https://doi.org/10.1016/0550-3213(77)90275-9},
}

@article{budzanowski_2010,
  title = {High Resolution Study of the {{$\Lambda$p}} Final State Interaction in the Reaction p+p\textrightarrow{{K}}++({{$\Lambda$p}})},
  author = {Budzanowski, A. and  others},
  collaboration = {HIRES Collaboration},
  year = {2010},
  month = apr,
  journal = {Physics Letters B},
  volume = {687},
  number = {1},
  pages = {31--35},
  issn = {0370-2693},
  urldate = {2023-06-20},
  langid = {english},
  doi = {10.1016/j.physletb.2010.02.082},
  url = {https://doi.org/10.1016/j.physletb.2010.02.082},
}

@article{cline_1968,
  title = {{$\Lambda$} p {{Interaction Near $\Sigma$ N Threshold}}},
  author = {Cline, D. and Laumann, R. and Mapp, J.},
  year = {1968},
  month = jun,
  journal = {Phys. Rev. Lett.},
  volume = {20},
  number = {25},
  pages = {1452--1455},
  issn = {0031-9007},
  urldate = {2023-05-15},
  langid = {english},
  doi = {10.1103/PhysRevLett.20.1452},
  url = {https://doi.org/10.1103/PhysRevLett.20.1452},
}

@article{dalitz_1981,
  title = {{$\Lambda$}- and {$\sum$}-Hypernuclear Physics},
  author = {Dalitz, R. H.},
  year = {1981},
  month = feb,
  journal = {Nuclear Physics A},
  volume = {354},
  number = {1},
  pages = {101--126},
  issn = {0375-9474},
  urldate = {2023-07-07},
  langid = {english},
  doi = {10.1016/0375-9474(81)90595-9},
  url = {https://doi.org/10.1016/0375-9474(81)90595-9},
}

@article{eisele_1971,
  title = {Elastic Σ±p scattering at low energies},
  journal = {Physics Letters B},
  volume = {37},
  number = {2},
  pages = {204-206},
  year = {1971},
  issn = {0370-2693},
  doi = {10.1016/0370-2693(71)90053-0},
  url = {https://www.sciencedirect.com/science/article/pii/0370269371900530},
  author = {F. Eisele and H. Filthuth and W. Föhlisch and V. Hepp and G. Zech}
}

@article{engenmann_1966,
  title = {Inelastic Σ−p-interactions at low momenta},
  journal = {Physics Letters},
  volume = {21},
  number = {5},
  pages = {587-589},
  year = {1966},
  issn = {0031-9163},
  doi = {10.1016/0031-9163(66)91310-2},
  url = {https://www.sciencedirect.com/science/article/pii/0031916366913102},
  author = {R. Engelmann and H. Filthuth and V. Hepp and E. Kluge}
}

@article{dalitz_1982,
  title = {{{TheK}}-d\textrightarrow p{$\lambda\pi-$} Reaction},
  author = {Dalitz, R. H. and Deloff, A.},
  year = {1982},
  month = sep,
  journal = {Czech J Phys},
  volume = {32},
  number = {9},
  pages = {1021--1039},
  issn = {1572-9486},
  urldate = {2023-07-07},
  langid = {english},
  doi = {10.1007/BF01597174},
  url = {https://doi.org/10.1007/BF01597174},
}

@article{eastwood_1971,
  title = {2130-{{MeV}} \$\{\textbackslash ensuremath\{\textbackslash\vphantom{\}\}}{{Lambda}}\vphantom\{\}\vphantom\{\}\^\{0\}p\$ {{Mass Enhancement}} in the {{Reaction}} \$\{\vphantom\}{{K}}\vphantom\{\}\^\{\textbackslash ensuremath\{-\}\}d\textbackslash ensuremath\{\textbackslash rightarrow\}\{\textbackslash ensuremath\{\textbackslash\vphantom{\}\}}{{Lambda}}\vphantom\{\}\vphantom\{\}\^\{0\}p\{\textbackslash ensuremath\{\textbackslash pi\}\}\^\{\textbackslash ensuremath\{-\}\}\$ at 1.45 and 1.65 \$\textbackslash frac\{\textbackslash mathrm\{\vphantom{\}\}}{{GeV}}\vphantom\{\}\vphantom\{\}\{c\}\$},
  author = {Eastwood, D. and Fry, J. R. and Heathcote, F. R. and Islam, G. S. and Candlin, D. J. and Copley, G. and Evans, G. R. and Campbell, J. R. and Morton, W. T. and Negus, P. J. and Counihan, M. J. and Goyal, D. P. and Miller, D. B. and Schwarzschild, B.},
  year = {1971},
  month = jun,
  journal = {Phys. Rev. D},
  volume = {3},
  number = {11},
  pages = {2603--2606},
  publisher = {{American Physical Society}},
  urldate = {2023-05-15},
  doi = {10.1103/PhysRevD.3.2603},
  url = {https://doi.org/10.1103/PhysRevD.3.2603},
}

@article{fujiwara_1996,
  title = {Effective Meson-Exchange Potentials in the \$\{\textbackslash mathrm\{\vphantom{\}\}}{{SU}}\vphantom\{\}\vphantom\{\}\_\{6\}\$ Quark Model for {{NN}} and {{YN}} Interactions},
  author = {Fujiwara, Y. and Nakamoto, C. and Suzuki, Y.},
  year = {1996},
  month = nov,
  journal = {Phys. Rev. C},
  volume = {54},
  number = {5},
  pages = {2180--2200},
  publisher = {{American Physical Society}},
  urldate = {2023-06-20},
  doi = {10.1103/PhysRevC.54.2180},
  url = {https://doi.org/10.1103/PhysRevC.54.2180},
}

@article{fujiwara_2007,
  title = {Baryon\textendash Baryon Interactions in the {{SU6}} Quark Model and Their Applications to Light Nuclear Systems},
  author = {Fujiwara, Y. and Suzuki, Y. and Nakamoto, C.},
  year = {2007},
  month = apr,
  journal = {Progress in Particle and Nuclear Physics},
  volume = {58},
  number = {2},
  pages = {439--520},
  issn = {0146-6410},
  urldate = {2023-06-20},
  langid = {english},
  doi = {10.1016/j.ppnp.2006.08.001},
  url = {https://doi.org/10.1016/j.ppnp.2006.08.001},
}

@article{garcilazo_2007,
  title = {\$\textbackslash ensuremath\{\textbackslash\vphantom\}{{Lambda}}\vphantom\{\}\textbackslash mathit\{\vphantom\}{{NN}}\vphantom\{\}\$ and \$\textbackslash ensuremath\{\textbackslash\vphantom\}{{Sigma}}\vphantom\{\}\textbackslash mathit\{\vphantom\}{{NN}}\vphantom\{\}\$ Systems at Threshold},
  author = {Garcilazo, H. and {Fern{\'a}ndez-Caram{\'e}s}, T. and Valcarce, A.},
  year = {2007},
  month = mar,
  journal = {Phys. Rev. C},
  volume = {75},
  number = {3},
  pages = {034002},
  publisher = {{American Physical Society}},
  urldate = {2023-06-20},
  doi = {10.1103/PhysRevC.75.034002},
  url = {https://doi.org/10.1103/PhysRevC.75.034002},
}

@article{haidenbauer_2005,
  title = {J\textbackslash "ulich Hyperon-Nucleon Model Revisited},
  author = {Haidenbauer, J. and Mei{\ss}ner, Ulf-G.},
  year = {2005},
  month = oct,
  journal = {Phys. Rev. C},
  volume = {72},
  number = {4},
  pages = {044005},
  publisher = {{American Physical Society}},
  urldate = {2023-04-24},
  doi = {10.1103/PhysRevC.72.044005},
  url = {https://doi.org/10.1103/PhysRevC.72.044005},
}

@article{haidenbauer_2013,
  title = {Hyperon\textendash Nucleon Interaction at next-to-Leading Order in Chiral Effective Field Theory},
  author = {Haidenbauer, J. and Petschauer, S. and Kaiser, N. and Mei{\ss}ner, U. -G. and Nogga, A. and Weise, W.},
  year = {2013},
  month = oct,
  journal = {Nuclear Physics A},
  volume = {915},
  pages = {24--58},
  issn = {0375-9474},
  urldate = {2023-06-20},
  langid = {english},
  doi = {10.1016/j.nuclphysa.2013.06.008},
  url = {https://doi.org/10.1016/j.nuclphysa.2013.06.008},
}

@article{haidenbauer_2020,
  title = {Hyperon\textendash Nucleon Interaction within Chiral Effective Field Theory Revisited},
  author = {Haidenbauer, J. and Mei{\ss}ner, U.-G. and Nogga, A.},
  year = {2020},
  month = mar,
  journal = {Eur. Phys. J. A},
  volume = {56},
  number = {3},
  pages = {91},
  issn = {1434-601X},
  urldate = {2023-06-20},
  langid = {english},
  doi = {10.1140/epja/s10050-020-00100-4},
  url = {https://doi.org/10.1140/epja/s10050-020-00100-4},
}

@article{haidenbauer_2021,
  title = {Constraints on the \$\$\textbackslash{{varLambda}} \$\$-{{Neutron Interaction}} from {{Charge Symmetry Breaking}} in the \$\$\textbackslash mathbf \{\^4\_\textbackslash\vphantom\}{{Lambda}} \textbackslash mathrm\{\vphantom\}{{He}}\vphantom\{\}\vphantom\{\}\$\$- \$\$\textbackslash mathbf \{\^4\_\textbackslash\vphantom\}{{Lambda}} \textbackslash mathrm\{\vphantom\}{{H}}\vphantom\{\}\vphantom\{\}\$\${{Hypernuclei}}},
  author = {Haidenbauer, Johann and Mei{\ss}ner, Ulf-G. and Nogga, Andreas},
  year = {2021},
  month = oct,
  journal = {Few-Body Syst},
  volume = {62},
  number = {4},
  pages = {105},
  issn = {1432-5411},
  urldate = {2023-04-24},
  langid = {english},
  doi = {10.1007/s00601-021-01684-3},
  url = {https://doi.org/10.1007/s00601-021-01684-3},
}

@article{haidenbauer_2023,
  title = {Hyperon–nucleon interaction in chiral effective field theory at next-to-next-to-leading order},
  volume = {59},
  ISSN = {1434-601X},
  url = {https://doi.org/10.1140/epja/s10050-023-00960-6},
  number = {3},
  pages = {63},
  journal = {Eur. Phys. J. A},
  publisher = {Springer Science and Business Media LLC},
  author = {Haidenbauer,  Johann and Meißner,  Ulf-G. and Nogga,  Andreas and Le,  Hoai},
  year = {2023},
  month = mar,
  doi = {10.1140/epja/s10050-023-00960-6},
}

@article{hauptman_1977,
  title = {Experimental study of Λp and Ξ0p interactions in the range 1 – 10 GeV/c},
  journal = {Nuclear Physics B},
  volume = {125},
  number = {1},
  pages = {29-51},
  year = {1977},
  issn = {0550-3213},
  doi = {10.1016/0550-3213(77)90222-X},
  url = {https://www.sciencedirect.com/science/article/pii/055032137790222X},
  author = {J.M. Hauptman and J.A. Kadyk and G.H. Trilling},
}

@article{holzenkamp_1989,
  title = {A Meson Exchange Model for the Hyperon-Nucleon Interaction},
  author = {Holzenkamp, B. and Holinde, K. and Speth, J.},
  year = {1989},
  month = aug,
  journal = {Nuclear Physics A},
  volume = {500},
  number = {3},
  pages = {485--528},
  issn = {0375-9474},
  urldate = {2023-06-20},
  langid = {english},
  doi = {10.1016/0375-9474(89)90223-6},
  url = {https://doi.org/10.1016/0375-9474(89)90223-6},
}

@article{ichikawa_2022,
  title = {High Resolution Spectroscopy of the ``{{$\Sigma$N}} Cusp'' by Using the d({{K}}-, {$\Pi-$}) Reaction},
  author = {Ichikawa, Y. and others},
  collaboration = {J-PARC E90 Collaboration},
  year = {2022},
  journal = {EPJ Web Conf.},
  volume = {271},
  pages = {02012},
  publisher = {{EDP Sciences}},
  issn = {2100-014X},
  urldate = {2023-06-20},
  copyright = {\textcopyright{} The Authors, published by EDP Sciences, 2022},
  langid = {english},
  doi = {10.1051/epjconf/202227102012},
  url = {https://doi.org/10.1051/epjconf/202227102012},
}

@article{iizawa_2022,
  title = {\$\{\vphantom\}{{K}}\vphantom\{\}\^\{\textbackslash ensuremath\{-\}\}d\textbackslash ensuremath\{\textbackslash rightarrow\}\textbackslash ensuremath\{\textbackslash pi\}\textbackslash mathrm\{\textbackslash ensuremath\{\textbackslash\vphantom{\}\}}{{Lambda}}\vphantom\{\}\vphantom\{\}{{N}}\$ Reaction for Studying Charge Symmetry Breaking in the \$\textbackslash mathrm\{\textbackslash ensuremath\{\textbackslash\vphantom{\}\}}{{Lambda}}\vphantom\{\}\vphantom\{\}{{N}}\$ Interaction},
  author = {Iizawa, Yutaro and Jido, Daisuke and Ishikawa, Takatsugu},
  year = {2022},
  month = oct,
  journal = {Phys. Rev. C},
  volume = {106},
  number = {4},
  pages = {045201},
  publisher = {{American Physical Society}},
  urldate = {2023-01-24},
  doi = {10.1103/PhysRevC.106.045201},
  url = {https://doi.org/10.1103/PhysRevC.106.045201},
}

@article{jido_2009,
  title = {Kaonic Production of \$ \textbackslash{{Lambda}}\$(1405) off Deuteron Target in Chiral Dynamics},
  author = {Jido, D. and Oset, E. and Sekihara, T.},
  year = {2009},
  month = oct,
  journal = {Eur. Phys. J. A},
  volume = {42},
  number = {2},
  pages = {257},
  issn = {1434-601X},
  urldate = {2022-12-15},
  langid = {english},
  doi = {10.1140/epja/i2009-10875-5},
  url = {https://doi.org/10.1140/epja/i2009-10875-5},
}

@article{jido_2011,
  title = {Kaon-Induced \$ \textbackslash{{Lambda}}\$(1405) Production on a Deuteron Target at {{DAFNE}}},
  author = {Jido, D. and Oset, E. and Sekihara, T.},
  year = {2011},
  month = mar,
  journal = {Eur. Phys. J. A},
  volume = {47},
  number = {3},
  pages = {42},
  issn = {1434-601X},
  urldate = {2023-06-20},
  langid = {english},
  doi = {10.1140/epja/i2011-11042-3},
  url = {https://doi.org/10.1140/epja/i2011-11042-3},
}

@article{jido_2013,
  title = {The {{K}}-d \textrightarrow{} {{$\pi\Sigma$n}} Reaction Revisited},
  author = {Jido, Daisuke and Oset, Eulogio and Sekihara, Takayasu},
  year = {2013},
  month = aug,
  journal = {Eur. Phys. J. A},
  volume = {49},
  number = {8},
  pages = {95},
  issn = {1434-601X},
  urldate = {2023-06-16},
  langid = {english},
  doi = {10.1140/epja/i2013-13095-6},
  url = {https://doi.org/10.1140/epja/i2013-13095-6},
}

@article{kadyk_1971,
  title = {Λp interactions in momentum range 300 to 1500 MeV/c},
  journal = {Nuclear Physics B},
  volume = {27},
  number = {1},
  pages = {13-22},
  year = {1971},
  issn = {0550-3213},
  doi = {10.1016/0550-3213(71)90076-9},
  url = {https://www.sciencedirect.com/science/article/pii/0550321371900769},
  author = {J.A. Kadyk and G. Alexander and J.H. Chan and P. Gaposchkin and G.H. Trilling},
}

@article{kamano_2014,
  title = {Dynamical coupled-channels model of ${K}^{\ensuremath{-}}p$ reactions: Determination of partial-wave amplitudes},
  author = {Kamano, H. and Nakamura, S. X. and Lee, T.-S. H. and Sato, T.},
  journal = {Phys. Rev. C},
  volume = {90},
  issue = {6},
  pages = {065204},
  numpages = {40},
  year = {2014},
  month = {Dec},
  publisher = {American Physical Society},
  doi = {10.1103/PhysRevC.90.065204},
  url = {https://link.aps.org/doi/10.1103/PhysRevC.90.065204}
}

@article{karplus_1959,
  title = {Inelastic {{Final-State Interactions}}: \$\{\vphantom\}{{K}}\vphantom\{\}\^\{\textbackslash ensuremath\{-\}\}\$ {{Absorption}} in {{Deuterium}}},
  shorttitle = {Inelastic {{Final-State Interactions}}},
  author = {Karplus, Robert and Rodberg, Leonard S.},
  year = {1959},
  month = aug,
  journal = {Phys. Rev.},
  volume = {115},
  number = {4},
  pages = {1058--1069},
  publisher = {{American Physical Society}},
  urldate = {2023-07-07},
  doi = {10.1103/PhysRev.115.1058},
  url = {https://doi.org/10.1103/PhysRev.115.1058},
}

@article{kohno_2000,
  title = {Hyperon Single-Particle Potentials Calculated from {{SU6}} Quark-Model Baryon\textendash Baryon Interactions},
  author = {Kohno, M. and Fujiwara, Y. and Fujita, T. and Nakamoto, C. and Suzuki, Y.},
  year = {2000},
  month = jul,
  journal = {Nuclear Physics A},
  volume = {674},
  number = {1},
  pages = {229--245},
  issn = {0375-9474},
  urldate = {2023-06-20},
  langid = {english},
  doi = {10.1016/s0375-9474(00)00164-0},
  url = {https://doi.org/10.1016/s0375-9474(00)00164-0},
}

@article{korpa_2001,
  title = {Hyperon-Nucleon Scattering and Hyperon Masses in the Nuclear Medium},
  author = {Korpa, C. L. and Dieperink, A. E. L. and Timmermans, R. G. E.},
  year = {2001},
  month = dec,
  journal = {Phys. Rev. C},
  volume = {65},
  number = {1},
  pages = {015208},
  publisher = {{American Physical Society}},
  urldate = {2023-06-20},
  doi = {10.1103/PhysRevC.65.015208},
  url = {https://doi.org/10.1103/PhysRevC.65.015208},
}

@article{kotani_1959,
  title = {Analysis of {{K}}- Absorption in Deuterium},
  author = {Kotani, T. and Ross, M.},
  year = {1959},
  month = dec,
  journal = {Nuovo Cim},
  volume = {14},
  number = {6},
  pages = {1282--1309},
  issn = {1827-6121},
  urldate = {2023-07-07},
  langid = {english},
  doi = {10.1007/BF02962351},
  url = {https://doi.org/10.1007/BF02962351},
}

@article{li_2016,
  title = {Strangeness \${{S}}=\textbackslash ensuremath\{-\}1\$ Hyperon-Nucleon Scattering in Covariant Chiral Effective Field Theory},
  author = {Li, Kai-Wen and Ren, Xiu-Lei and Geng, Li-Sheng and Long, Bingwei},
  year = {2016},
  month = jul,
  journal = {Phys. Rev. D},
  volume = {94},
  number = {1},
  pages = {014029},
  publisher = {{American Physical Society}},
  urldate = {2023-06-20},
  doi = {10.1103/PhysRevD.94.014029},
  url = {https://doi.org/10.1103/PhysRevD.94.014029},
}

@article{li_2018,
  title = {Leading Order Relativistic Hyperon-Nucleon Interactions in Chiral Effective Field Theory*},
  author = {Li, Kai-Wen and Ren, Xiu-Lei and Geng, Li-Sheng and Long, Bing-Wei},
  year = {2018},
  month = jan,
  journal = {Chinese Phys. C},
  volume = {42},
  number = {1},
  pages = {014105},
  publisher = {{Chinese Physical Society and the Institute of High Energy Physics of the Chinese Academy of Sciences and the Institute of Modern Physics of the Chinese Academy of Sciences and IOP Publishing Ltd}},
  issn = {1674-1137},
  urldate = {2023-06-20},
  langid = {english},
  doi = {10.1088/1674-1137/42/1/014105},
  url = {https://doi.org/10.1088/1674-1137/42/1/014105},
}

@article{machleidt_2001,
  title = {High-Precision, Charge-Dependent {{Bonn}} Nucleon-Nucleon Potential},
  author = {Machleidt, R.},
  year = {2001},
  month = jan,
  journal = {Phys. Rev. C},
  volume = {63},
  number = {2},
  pages = {024001},
  publisher = {{American Physical Society}},
  urldate = {2023-04-24},
  doi = {10.1103/PhysRevC.63.024001},
  url = {https://doi.org/10.1103/PhysRevC.63.024001},
}

@article{miyagawa_2012,
  title = {Precise calculation of the two-step process for ${K}^{\ensuremath{-}}d\ensuremath{\rightarrow}\ensuremath{\pi}\ensuremath{\Sigma}n$ in the $\ensuremath{\Lambda}$(1405) resonance region},
  author = {Miyagawa, K. and Haidenbauer, J.},
  journal = {Phys. Rev. C},
  volume = {85},
  issue = {6},
  pages = {065201},
  numpages = {10},
  year = {2012},
  month = {Jun},
  publisher = {American Physical Society},
  doi = {10.1103/PhysRevC.85.065201},
  url = {https://link.aps.org/doi/10.1103/PhysRevC.85.065201}
}

@article{miyagawa_2018,
  title = {Faddeev approach to the reaction ${K}^{\ensuremath{-}}d\ensuremath{\rightarrow}\ensuremath{\pi}\mathrm{\ensuremath{\Sigma}}n$ at ${p}_{K}=1\phantom{\rule{0.16em}{0ex}}\mathrm{GeV}/c$},
  author = {Miyagawa, K. and Haidenbauer, J. and Kamada, H.},
  journal = {Phys. Rev. C},
  volume = {97},
  issue = {5},
  pages = {055209},
  numpages = {13},
  year = {2018},
  month = {May},
  publisher = {American Physical Society},
  doi = {10.1103/PhysRevC.97.055209},
  url = {https://link.aps.org/doi/10.1103/PhysRevC.97.055209}
}

@article{miwa_2021,
  title = {Measurement of the differential cross sections of the ${\mathrm{\ensuremath{\Sigma}}}^{\ensuremath{-}}p$ elastic scattering in momentum range 470 to 850 $\mathrm{MeV}/c$},
  author = {Miwa, K. and others},
  collaboration = {J-PARC E40 Collaboration},
  journal = {Phys. Rev. C},
  volume = {104},
  issue = {4},
  pages = {045204},
  numpages = {20},
  year = {2021},
  month = {Oct},
  publisher = {American Physical Society},
  doi = {10.1103/PhysRevC.104.045204},
  url = {https://link.aps.org/doi/10.1103/PhysRevC.104.045204}
}

@article{miwa_2022,
  title = {Precise Measurement of Differential Cross Sections of the ${\mathrm{\ensuremath{\Sigma}}}^{\ensuremath{-}}p\ensuremath{\rightarrow}\mathrm{\ensuremath{\Lambda}}n$ Reaction in Momentum Range $470--650\text{ }\text{ }\mathrm{MeV}/c$},
  author = {Miwa, K. and others},
  collaboration = {J-PARC E40 Collaboration},
  journal = {Phys. Rev. Lett.},
  volume = {128},
  issue = {7},
  pages = {072501},
  numpages = {6},
  year = {2022},
  month = {Feb},
  publisher = {American Physical Society},
  doi = {10.1103/PhysRevLett.128.072501},
  url = {https://link.aps.org/doi/10.1103/PhysRevLett.128.072501}
}

@article{nagels_2019,
  title = {Extended-Soft-Core Baryon-Baryon Model {{ESC16}}. {{II}}. {{Hyperon-nucleon}} Interactions},
  author = {Nagels, M. M. and Rijken, {\relax Th}. A. and Yamamoto, Y.},
  year = {2019},
  month = apr,
  journal = {Phys. Rev. C},
  volume = {99},
  number = {4},
  pages = {044003},
  publisher = {{American Physical Society}},
  urldate = {2023-06-20},
  doi = {10.1103/PhysRevC.99.044003},
  url = {https://doi.org/10.1103/PhysRevC.99.044003},
}

@article{ohnishi_2016,
  title = {Structure of the $\mathrm{\ensuremath{\Lambda}}(1405)$ and the ${K}^{\ensuremath{-}}d\ensuremath{\rightarrow}\ensuremath{\pi}\mathrm{\ensuremath{\Sigma}}n$ reaction},
  author = {Ohnishi, Shota and Ikeda, Yoichi and Hyodo, Tetsuo and Weise, Wolfram},
  journal = {Phys. Rev. C},
  volume = {93},
  issue = {2},
  pages = {025207},
  numpages = {12},
  year = {2016},
  month = {Feb},
  publisher = {American Physical Society},
  doi = {10.1103/PhysRevC.93.025207},
  url = {https://link.aps.org/doi/10.1103/PhysRevC.93.025207}
}

@article{petschauer_2020,
  title = {Hyperon-{{Nuclear Interactions From SU}}(3) {{Chiral Effective Field Theory}}},
  author = {Petschauer, Stefan and Haidenbauer, Johann and Kaiser, Norbert and Mei{\ss}ner, Ulf-G. and Weise, Wolfram},
  year = {2020},
  journal = {Frontiers in Physics},
  volume = {8},
  pages = {12},
  doi = {10.3389/fphy.2020.00012},
  issn = {2296-424X},
  urldate = {2023-06-20},
  url = {https://doi.org/10.3389/fphy.2020.00012},
}

@article{pigot_1985,
  title = {Study of the Production of a Strange ({{S}} = -1) Dibaryon in the Interactions of {{K}}- and {$\pi$}+ on Deuterium below 1.{{5GeV}}/c},
  author = {Pigot, C. and {de Brion}, J. P. and Caillet, A. and Cheze, J. B. and Derre, J. and Marel, G. and Pauli, E. and Vrana, G. and {d'Agostini}, G. and Auriemma, G. and Marini, G. and Martellotti, G. and Massa, F. and Nigro, A. and Rambaldi, A. and Sciubba, A. and Reucroft, S. and Roos, C. E. and Webster, M. and Williams, E. G. H.},
  year = {1985},
  month = jan,
  journal = {Nuclear Physics B},
  volume = {249},
  number = {1},
  pages = {172--188},
  issn = {0550-3213},
  urldate = {2023-06-20},
  langid = {english},
  doi = {10.1016/0550-3213(85)90045-8},
  url = {https://doi.org/10.1016/0550-3213(85)90045-8},
}

@article{polinder_2006,
  title = {Hyperon\textendash Nucleon Interactions\textemdash a Chiral Effective Field Theory Approach},
  author = {Polinder, Henk and Haidenbauer, Johann and Mei{\ss}ner, Ulf-G.},
  year = {2006},
  month = nov,
  journal = {Nuclear Physics A},
  volume = {779},
  pages = {244--266},
  issn = {0375-9474},
  urldate = {2023-06-20},
  langid = {english},
  doi = {10.1016/j.nuclphysa.2006.09.006},
  url = {https://doi.org/10.1016/j.nuclphysa.2006.09.006},
}

@article{ren_2020,
  title = {\$\textbackslash mathrm\{\textbackslash ensuremath\{\textbackslash\vphantom{\}\}}{{Lambda}}\vphantom\{\}\vphantom\{\}\$-Nucleon Scattering in Baryon Chiral Perturbation Theory},
  author = {Ren, X.-L. and Epelbaum, E. and Gegelia, J.},
  year = {2020},
  month = mar,
  journal = {Phys. Rev. C},
  volume = {101},
  number = {3},
  pages = {034001},
  publisher = {{American Physical Society}},
  urldate = {2023-06-20},
  doi = {10.1103/PhysRevC.101.034001},
  url = {https://doi.org/10.1103/PhysRevC.101.034001},
}

@article{reuber_1994,
  title = {Meson-Exchange Hyperon-Nucleon Interactions in Free Scattering and Nuclear Matter},
  author = {Reuber, A. and Holinde, K. and Speth, J.},
  year = {1994},
  month = apr,
  journal = {Nuclear Physics A},
  volume = {570},
  number = {3},
  pages = {543--579},
  issn = {0375-9474},
  urldate = {2023-06-20},
  langid = {english},
  doi = {10.1016/0375-9474(94)90073-6},
  url = {https://doi.org/10.1016/0375-9474(94)90073-6},
}

@article{rijken_1999,
  title = {Soft-Core Hyperon-Nucleon Potentials},
  author = {Rijken, {\relax Th}. A. and Stoks, V. G. J. and Yamamoto, Y.},
  year = {1999},
  month = jan,
  journal = {Phys. Rev. C},
  volume = {59},
  number = {1},
  pages = {21--40},
  publisher = {{American Physical Society}},
  urldate = {2023-04-24},
  doi = {10.1103/PhysRevC.59.21},
  url = {https://doi.org/10.1103/PhysRevC.59.21},
}

@article{rijken_2010,
  title = {Baryon-{{Baryon Interactions}}: \textemdash{} {{Nijmegen Extended-Soft-Core Models}} \textemdash},
  shorttitle = {Baryon-{{Baryon Interactions}}},
  author = {Rijken, Thomas A. and Nagels, M. M. and Yamamoto, Yasuo},
  year = {2010},
  month = jun,
  journal = {Progress of Theoretical Physics Supplement},
  volume = {185},
  pages = {14--71},
  issn = {0375-9687},
  urldate = {2023-06-20},
  doi = {10.1143/ptps.185.14},
  url = {https://doi.org/10.1143/ptps.185.14},
}

@article{rowley_2021,
  title = {Improved $\mathrm{\ensuremath{\Lambda}}p$ Elastic Scattering Cross Sections between 0.9 and $2.0\text{ }\text{ }\mathrm{GeV}/c$ as a Main Ingredient of the Neutron Star Equation of State},
  author = {Rowley, J. and others},
  collaboration = {CLAS Collaboration},
  journal = {Phys. Rev. Lett.},
  volume = {127},
  issue = {27},
  pages = {272303},
  numpages = {7},
  year = {2021},
  month = {Dec},
  publisher = {American Physical Society},
  doi = {10.1103/PhysRevLett.127.272303},
  url = {https://link.aps.org/doi/10.1103/PhysRevLett.127.272303}
}

@article{savage_1996,
  title = {Hyperon Masses in Nuclear Matter},
  author = {Savage, Martin J. and Wise, Mark B.},
  year = {1996},
  month = jan,
  journal = {Phys. Rev. D},
  volume = {53},
  number = {1},
  pages = {349--354},
  publisher = {{American Physical Society}},
  urldate = {2023-06-20},
  doi = {10.1103/PhysRevD.53.349},
  url = {https://doi.org/10.1103/PhysRevD.53.349},
}

@article{sechi-zorn_1968,
  title = {Low-Energy $\ensuremath{\Lambda}$-Proton Elastic Scattering},
  author = {Sechi-Zorn, B. and Kehoe, B. and Twitty, J. and Burnstein, R. A.},
  journal = {Phys. Rev.},
  volume = {175},
  issue = {5},
  pages = {1735--1740},
  numpages = {0},
  year = {1968},
  month = {Nov},
  publisher = {American Physical Society},
  doi = {10.1103/PhysRev.175.1735},
  url = {https://link.aps.org/doi/10.1103/PhysRev.175.1735}
}

@article{sims_1971,
  title = {Study of the Nonimpulse Events in the Reactions ${K}^{\ensuremath{-}}d\ensuremath{\rightarrow}\ensuremath{\Lambda}\ensuremath{\pi}p$ and $\ensuremath{\Lambda}\ensuremath{\pi}\ensuremath{\pi}N$ at $670\ensuremath{-}925 \frac{\mathrm{GeV}}{c}$},
  author = {Sims, W. H. and O'Neall, J. S. and Albright, J. R. and Brucker, E. B. and Lannutti, J. E.},
  journal = {Phys. Rev. D},
  volume = {3},
  issue = {5},
  pages = {1162--1164},
  numpages = {0},
  year = {1971},
  month = {Mar},
  publisher = {American Physical Society},
  doi = {10.1103/PhysRevD.3.1162},
  url = {https://link.aps.org/doi/10.1103/PhysRevD.3.1162}
}

@article{song_2018,
  title = {Strangeness \${{S}}=\textbackslash ensuremath\{-\}1\$ Hyperon-Nucleon Interactions: {{Chiral}} Effective Field Theory versus Lattice {{QCD}}},
  shorttitle = {Strangeness \${{S}}=\textbackslash ensuremath\{-\}1\$ Hyperon-Nucleon Interactions},
  author = {Song, Jing and Li, Kai-Wen and Geng, Li-Sheng},
  year = {2018},
  month = jun,
  journal = {Phys. Rev. C},
  volume = {97},
  number = {6},
  pages = {065201},
  publisher = {{American Physical Society}},
  urldate = {2023-06-20},
  doi = {10.1103/PhysRevC.97.065201},
  url = {https://doi.org/10.1103/PhysRevC.97.065201},
}

@article{toker_1979,
  title = {The {{K}}-d \textrightarrow{} {$\pi-$}{{$\Lambda$p}} Reaction at Low Energies in the Faddeev Formalism},
  author = {Toker, G. and Gal, A. and Eisenberg, J. M.},
  year = {1979},
  month = dec,
  journal = {Physics Letters B},
  volume = {88},
  number = {3},
  pages = {235--238},
  issn = {0370-2693},
  urldate = {2023-07-07},
  langid = {english},
  doi = {10.1016/0370-2693(79)90455-6},
  url = {https://doi.org/10.1016/0370-2693(79)90455-6},
}

@article{toker_1981,
  title = {The {{YN}} Interactions and {{K}}- Reactions on Deuterium at Low Energies},
  author = {Toker, G. and Gal, A. and Eisenberg, J. M.},
  year = {1981},
  month = jun,
  journal = {Nuclear Physics A},
  volume = {362},
  number = {2},
  pages = {405--430},
  issn = {0375-9474},
  urldate = {2023-07-07},
  langid = {english},
  doi = {10.1016/0375-9474(81)90502-9},
  url = {https://doi.org/10.1016/0375-9474(81)90502-9},
}

@article{tominaga_1998,
  title = {A One-Boson-Exchange Potential for {{$\Lambda$N}}, {{$\Lambda\Lambda$}} and {{$\Xi$N}} Systems and Hypernuclei},
  author = {Tominaga, K. and Ueda, T. and Yamaguchi, M. and Kijima, N. and Okamoto, D. and Miyagawa, K. and Yamada, T.},
  year = {1998},
  month = nov,
  journal = {Nuclear Physics A},
  volume = {642},
  number = {3},
  pages = {483--505},
  issn = {0375-9474},
  urldate = {2023-06-20},
  langid = {english},
  doi = {10.1016/s0375-9474(98)00485-0},
  url = {https://doi.org/10.1016/s0375-9474(98)00485-0},
}

@article{tominaga_2001,
  title = {Effective One-Boson-Exchange Potential for {{$\Lambda$N}} and {{$\Sigma$N}} Systems and Hypertriton},
  author = {Tominaga, K. and Ueda, T.},
  year = {2001},
  month = oct,
  journal = {Nuclear Physics A},
  volume = {693},
  number = {3},
  pages = {731--754},
  issn = {0375-9474},
  urldate = {2023-06-20},
  langid = {english},
  doi = {10.1016/s0375-9474(01)00882-x},
  url = {https://doi.org/10.1016/s0375-9474(01)00882-x},
}

@article{torres_1986,
  title = {{{K}}- Absorption Reactions from Rest in Deuterium},
  author = {Torres, M. and Dalitz, R. H. and Deloff, A.},
  year = {1986},
  month = jul,
  journal = {Physics Letters B},
  volume = {174},
  number = {2},
  pages = {213--218},
  issn = {0370-2693},
  urldate = {2023-07-07},
  langid = {english},
  doi = {10.1016/0370-2693(86)90744-6},
  url = {https://doi.org/10.1016/0370-2693(86)90744-6},
}

@article{workman_2012,
  title = {Parameterization dependence of $T$-matrix poles and eigenphases from a fit to $\ensuremath{\pi}N$ elastic scattering data},
  author = {Workman, R. L. and Arndt, R. A. and Briscoe, W. J. and Paris, M. W. and Strakovsky, I. I.},
  journal = {Phys. Rev. C},
  volume = {86},
  issue = {3},
  pages = {035202},
  numpages = {6},
  year = {2012},
  month = {Sep},
  publisher = {American Physical Society},
  doi = {10.1103/PhysRevC.86.035202},
  url = {https://link.aps.org/doi/10.1103/PhysRevC.86.035202}
}

@article{yamagata-sekihara_2013,
  title = {Production of Hyperon Resonances Induced by Kaons on a Deuteron Target},
  author = {{Yamagata-Sekihara}, J. and Sekihara, T. and Jido, D.},
  year = {2013},
  month = apr,
  journal = {Progress of Theoretical and Experimental Physics},
  volume = {2013},
  number = {4},
  pages = {043D02},
  issn = {2050-3911},
  urldate = {2023-01-24},
  doi = {10.1093/ptep/ptt009},
  url = {https://doi.org/10.1093/ptep/ptt009},
}

@article{yasunaga_2025,
  title = {${K}^{\ensuremath{-}}d\ensuremath{\rightarrow}\ensuremath{\pi}\mathrm{\ensuremath{\Lambda}}N$ reaction with in-flight kaons for studying the $\mathrm{\ensuremath{\Lambda}}N$ interaction},
  author = {Yasunaga, Shunsuke and Jido, Daisuke and Ishikawa, Takatsugu},
  journal = {Phys. Rev. C},
  volume = {112},
  issue = {1},
  pages = {015208},
  numpages = {15},
  year = {2025},
  month = {Jul},
  publisher = {American Physical Society},
  doi = {10.1103/8d48-w2dr},
  url = {https://link.aps.org/doi/10.1103/8d48-w2dr}
}

@article{yasunaga_2025a,
  author = {S. Yasunaga and D. Jido},
  title = {Theoretical study of the ${\ensuremath{\Sigma}}N$ cusp in the ${K}^{\ensuremath{-}}d\ensuremath{\rightarrow}\ensuremath{\pi}\mathrm{\ensuremath{\Lambda}}N$ reaction},
  journal = {PoS},
  volume = {QNP2024},
  pages = {101},
  year = {2025},
  doi = {10.22323/1.465.0101},
  url = {https://doi.org/10.22323/1.465.0101}
}

@article{yasunaga_2026,
  author = {Yasunaga, Shunsuke and Jido, Daisuke},
  title = {Hyperon--nucleon interaction through the {$K^-d\to\pi\Lambda N$} reaction},
  journal = {Journal of Subatomic Particles and Cosmology},
  volume = {6},
  pages = {100462},
  year = {2026},
  doi = {10.1016/j.jspc.2026.100462},
  url = {https://www.sciencedirect.com/science/article/pii/S3050480526001780}
}

@misc{sone_2026,
  author        = {Sone, Katsuyoshi and Hyodo, Tetsuo},
  year          = {2026},
  eprint        = {2605.26658},
  archivePrefix = {arXiv},
  primaryClass  = {hep-ph},
  doi           = {10.48550/arXiv.2605.26658},
  url           = {https://arxiv.org/abs/2605.26658}
}

@article{lacombe_1980,
  author = {Lacombe, M. and Loiseau, B. and Richard, J. M. and Vinh Mau, R. and C{\^o}t{\'e}, J. and Pir{\`e}s, P. and de Tourreil, R.},
  title = {Parametrization of the Paris potential},
  journal = {Phys. Rev. C},
  volume = {21},
  pages = {861--873},
  year = {1980},
  doi = {10.1103/PhysRevC.21.861},
  url = {https://doi.org/10.1103/PhysRevC.21.861}
}

@article{wiringa_1995,
  author = {Wiringa, R. B. and Stoks, V. G. J. and Schiavilla, R.},
  title = {Accurate nucleon-nucleon potential with charge-independence breaking},
  journal = {Phys. Rev. C},
  volume = {51},
  pages = {38--51},
  year = {1995},
  doi = {10.1103/PhysRevC.51.38},
  url = {https://doi.org/10.1103/PhysRevC.51.38}
}

@article{inoue_2010,
  author        = {Inoue, Takashi and Ishii, Noriyoshi and Aoki, Sinya
                   and Doi, Takumi and Hatsuda, Tetsuo and Ikeda, Yoichi
                   and Murano, Keiko and Nemura, Hidekatsu and Sasaki, Kenji},
  collaboration = {HAL QCD Collaboration},
  title         = {{Baryon-Baryon Interactions in the Flavor SU(3) Limit
                    from Full QCD Simulations on the Lattice}},
  journal       = {Prog. Theor. Phys.},
  volume        = {124},
  pages         = {591--603},
  year          = {2010},
  doi           = {10.1143/PTP.124.591},
  url = {https://doi.org/10.1143/PTP.124.591},
}

@misc{liu_2026,
  author        = {Liu, Hang and Liu, Liuming and Tan, Jin-Xin
                   and Wang, Wei and Yan, Haobo and Zhu, Qian-Teng},
  year          = {2026},
  eprint        = {2603.05854},
  archivePrefix = {arXiv},
  primaryClass  = {hep-lat},
  doi           = {10.48550/arXiv.2603.05854},
  url = {https://arxiv.org/abs/2603.05854},
}

@article{machner_2013,
    author = {Machner, H. and Haidenbauer, J. and Hinterberger, F. and Magiera, A. and Niskanen, J. A. and Ritman, J. and Siudak, R.},
    title = {Study of the {$\Lambda p$} Interaction Close to the {$\Sigma^+ n$} and {$\Sigma^0 p$} Thresholds},
    journal = {Nucl. Phys. A},
    volume = {901},
    pages = {65--88},
    year = {2013},
    doi = {10.1016/j.nuclphysa.2013.01.031},
  url = {https://doi.org/10.1016/j.nuclphysa.2013.01.031},
}
\end{document}